\documentclass[12pt,a4paper]{article}

\usepackage[T1]{fontenc}
\usepackage[utf8]{inputenc}
\usepackage{mathptmx}                 
\usepackage[a4paper,margin=1in]{geometry}
\usepackage{setspace}
\usepackage{microtype}

\usepackage{amsmath}
\usepackage{amssymb}
\usepackage{bm}
\usepackage{graphicx}
\usepackage{booktabs}
\usepackage{multirow}
\usepackage{url}
\usepackage{xcolor}
\usepackage{algorithm}
\usepackage{algorithmicx}
\usepackage{algpseudocode}
\newcommand{\pending}[1]{--}

\usepackage[round,authoryear]{natbib} 
\usepackage[hidelinks]{hyperref}      

\renewenvironment{abstract}{%
  \small
  {\noindent\bfseries \abstractname\par}\smallskip
  \noindent\ignorespaces}{\par\medskip}

\begin{document}

\title{\Large Wavelet-Based Bayesian Hierarchical Modeling with Regularized Horseshoe Priors for Spatially Correlated Functional Data} 

\author{%
\small  Alvaro Alexander Burbano-Moreno\textsuperscript{a}\thanks{\textsuperscript{*}Corresponding author.\indent \textit{Email address:} \texttt{aamoreno@unicamp.br}}, \quad
  Alex Rodrigo dos Santos Sousa\textsuperscript{a}, \quad
  Luiz Koodi Hotta\textsuperscript{a} \\
  \small \textsuperscript{a}Instituto de Matemática, Estatística e Computação Científica. Universidade Estadual de Campinas\\
  \small (UNICAMP), Campinas, SP, Brasil
}

\date{}
\maketitle
\begin{abstract}
Environmental monitoring networks increasingly record pollutant levels as curves observed over time at fixed stations. Yet, exposure and risk assessment often require predicting these curves, with their uncertainty, at unmonitored locations. Conventional smooth-basis methods for such spatially correlated functional data, including spline-based kriging, tend to over-smooth acute, high-frequency episodes such as pollution peaks. We propose a parsimonious wavelet-based Bayesian hierarchical model that anchors a single, shared Matérn correlation matrix at each resolution level, avoiding the parameter inflation of coefficient-specific spatial processes. Adaptive sparsity is enforced through a spatially informed regularized horseshoe prior, allowing the model to borrow strength across neighboring locations. At the same time, a non-centered parameterization ensures stable inference under the No-U-Turn Sampler. In simulation studies across distinct signal-to-noise regimes, incorporating spatial dependence substantially stabilizes reconstruction under severe noise relative to strictly independent models. Applied to PM$_{10}$ concentrations from Mexico City and benchmarked against ordinary kriging for functional data and a hierarchical Bayesian wavelet alternative, the model captures acute pollution events and, beyond improving point prediction, delivers sharper and better-calibrated credible intervals at unobserved locations.
\end{abstract}


\bigskip

\section{Introduction}\label{sec:Introduction}
The rapid advancement of automated monitoring networks and sensor technologies has fundamentally transformed the collection and analysis of environmental, epidemiological, and physical datasets. Measurements previously treated as discrete longitudinal vectors are now often analyzed as continuous functional curves over specific domains. This methodological shift has driven significant progress in Functional Data Analysis (FDA) \citep{ramsay2005,ferraty2006,kokoszka2017}. Although early FDA methodologies primarily assumed statistical independence among sampled curves, contemporary applications often involve data that violate this assumption. In domains such as air-quality monitoring, regional meteorological studies, and epidemiological tracking, functional observations are intrinsically associated with geographic coordinates \citep{cressie1993, banerjee:2014}. The complex interplay between functional dynamics and geographic proximity requires methodological frameworks that can model spatially correlated functional data \citep{mateu2021}. This requirement is particularly pronounced in air-quality monitoring, where networks record pollutant concentrations at fixed stations, yet decisions regarding exposure and health risk frequently pertain to unmonitored locations. Consequently, predicting complete pollutant curves at unobserved sites, along with a rigorous quantification of their associated uncertainty, constitutes a central inferential objective.

Traditional spatial functional methodologies, such as Ordinary Kriging for Functional Data (OKFD) \citep{giraldo:2011}, have primarily relied on basis expansion techniques, including B-splines and Fourier series, to address the infinite-dimensional nature of curves. Although these conventional bases are effective for modeling smooth, global trends, they often fail to capture non-stationary features or sudden localized spikes. In particular, spline-based spatial models tend to over-smooth functional trajectories, systematically underestimating sharp local variations. In environmental monitoring, such localized features often correspond to acute episodes, such as sudden pollution peaks, whose accurate reconstruction is directly relevant to exposure assessment and risk evaluation. To overcome these limitations, wavelets \citep{daubechies:1992, vidakovic:1999, vidakovic:2001, Mallat:2008} provide a robust alternative. Owing to their multiresolution properties, wavelets can represent both smooth macroscopic characteristics and localized irregularities using a relatively small number of non-zero coefficients.

However, applying the discrete wavelet transform results in a high-dimensional parameter space, necessitating robust sparsity-inducing techniques. In Bayesian wavelet modeling, sparsity has typically been enforced through discrete spike-and-slab priors \citep{zhou:2021, vannucci:2021}. Although theoretically grounded, these priors can present significant computational challenges when exploring complex posterior spaces, particularly for gradient-based sampling algorithms such as Hamiltonian Monte Carlo (HMC) and the No-U-Turn Sampler (NUTS). To maintain theoretical identifiability while ensuring computational feasibility, level-dependent continuous shrinkage strategies provide a highly tractable alternative. Among these approaches, the Regularized Horseshoe (RHS) prior \citep{piironen:2017} has been extensively studied and applied with strong results across diverse statistical contexts, owing to its robust shrinkage and adaptive selection properties \citep{uribe:2023, chen:2025, Kun:2026}.

The intersection of FDA, geostatistics, and wavelet methodologies has yielded significant advancements that inform the present study. Early research by \citet{morris:2006} introduced wavelet-based functional mixed models, offering a unified framework for modeling complex, irregular functional data by projecting observed profiles into the wavelet domain to flexibly accommodate covariance structures. Building on this, \citet{fernandez:2015} proposed moment regression and Bayesian wavelet techniques tailored for spatially correlated functional data. More recently, \citet{song:2019} developed a hierarchical Bayesian framework for predicting spatially correlated curves in the wavelet domain. This framework assigns distinct spatial covariance matrices, each with its own variance and decay parameters, to the wavelet coefficients across all multiresolution levels and translations. Although this approach provides considerable flexibility, it requires estimating hundreds of spatial hyperparameters, each informed by the number of monitoring sites, resulting in substantial computational demands and identifiability challenges for the spatial decay and variance parameters. Anticipating these issues, \citet{song:2019} suggested employing level-wise spatial process priors for the coefficients to reduce the number of model parameters and alleviate computational complexity. The present study adopts and extends this recommendation beyond a simple level-wise reduction. Bayesian approaches to spatially correlated functional data have also been developed outside the wavelet domain, using alternative basis systems and dependence structures \citep{baladandayuthapani:2008, zhang:2016}.

Continuous global-local shrinkage methods have also been adapted to incorporate spatial structure, but typically in contexts distinct from the present study. For example, spatial horseshoe priors have been proposed for signal detection on regular image grids \citep{jhuang:2019}, and spatially dependent global-local priors have been developed for areal selection under conditional autoregressive dependence \citep{zhu:2025}. Horseshoe-type shrinkage has further been examined in the wavelet domain for sparse sequence problems \citep{zhai:2026}. These approaches operate on lattice or areal supports and do not address curve prediction at unobserved coordinates. To our knowledge, these methodological developments have not yet been integrated for spatially correlated functional data with point-referenced support. Specifically, the RHS prior remains largely unexplored in combination with wavelet-based multiresolution representations and an explicit geostatistical correlation structure over the continuous domain of curve-generating locations, which is essential for spatial prediction at unmonitored sites. This gap is not simply a matter of unexplored methodology; coupling continuous shrinkage with an explicit spatial process requires a modeling choice that the integration renders unavoidable. When the spatial process is specified through a full covariance matrix, its marginal variance and the local shrinkage penalty both scale the same coefficients, rendering them non-identifiable. This confounding differs from the decay-versus-variance identifiability issues previously discussed for \citet{song:2019}; it arises specifically from the interaction between continuous shrinkage and spatial variance, an interaction that, to our knowledge, has not been made explicit in the wavelet RHS context. We accordingly anchor spatial dependence through a Matérn correlation matrix with unit marginal variance, a standard identification strategy in hierarchical spatial models, so that the level-wise shrinkage scale alone governs the amplitude of the coefficients (Section~\ref{sec:rhs}).

To address this gap, this study introduces a parsimonious Wavelet-Based Bayesian Hierarchical Model that adapts the RHS prior for spatially correlated functional data. The approach integrates Bayesian hierarchical modeling, wavelet basis decomposition, and a streamlined spatial dependence structure. Rather than estimating independent spatial processes for thousands of coefficients, the proposed framework anchors a single, shared Matérn spatial process at each resolution level, embedding spatial dependence directly into the RHS regularization mechanism. This integration enables adaptive thresholding that suppresses noise while preserving genuine high-frequency features by borrowing strength across neighboring locations, rather than acting on each site in isolation. Consequently, even when local data are heavily affected by observational noise, information from neighboring locations guides the continuous shrinkage. Although each component is well established individually, their combination yields this behavior. Two design choices make this construction viable. First, a single shared Matérn correlation, rather than a full covariance, is anchored at each resolution level, in contrast to the per-coefficient spatial processes of \citet{song:2019}, thereby removing the confounding between spatial scale and local shrinkage penalty (Section~\ref{sec:rhs}). Second, we formulate the full hierarchy using a strictly non-centered parameterization, which mitigates the funnel-shaped posterior geometries that hinder gradient-based sampling of horseshoe-type priors and supports stable inference under the NUTS algorithm. Additionally, prediction at unobserved sites reduces to a closed-form Gaussian conditioning step in the wavelet domain, enabling recovery of complete functional trajectories with well-calibrated credible intervals without requiring additional posterior simulation (Section~\ref{sec:prediction}).

The proposal is evaluated through a comprehensive simulation study and a real-data application against two baselines: the classical OKFD method \citep{giraldo:2011} and the hierarchical Bayesian wavelet model of \citet{song:2019}, the latter re-implemented within the same \texttt{Stan} environment so that observed differences reflect the underlying statistical architectures rather than implementation details. Applied to daily PM$_{10}$ concentrations from Mexico City's monitoring network, the proposed methodology outperforms OKFD in point prediction while providing the calibrated predictive intervals that kriging cannot offer, and it yields sharper, better-calibrated intervals than the \citet{song:2019} benchmark. The remainder of this paper is organized as follows. Section~\ref{sec:2} presents the theoretical framework, detailing the wavelet background, the hierarchical formulation of the spatial Regularized Horseshoe prior, and the closed-form scheme for spatial prediction. Section~\ref{sec:3} reports the simulation study, organized around three objectives: recovering signals across three SNR regimes, quantifying the gain from borrowing spatial strength relative to a strictly independent baseline, and evaluating predictive performance at unobserved locations. Section~\ref{sec:realdata} applies the methodology to PM$_{10}$ concentrations in Mexico City, and Section~\ref{sec:5} concludes with a discussion of the main findings and future directions.
\section{Theoretical Framework}\label{sec:2}
\subsection{Background on Wavelets}\label{sec:2.1}
Wavelets are basis functions that enable efficient approximation of other functions using a relatively small number of non-zero wavelet coefficients. The construction of a family of wavelets relies on two carefully chosen orthonormal basis functions: the scaling function \( \phi \) and the mother wavelet \( \psi \). From these functions, a system of wavelets is generated through dilation and translation operations applied to \( \phi \) and \( \psi \), yielding the following expressions:
\(\phi_{l,k}(t) = 2^{l/2} \, \phi(2^l t - k)\) and \(\psi_{l,k}(t) = 2^{l/2} \, \psi(2^l t - k)\), 
where \( l \in \mathbb{Z} \) denotes the scale level (or resolution), and \( k \in \mathbb{Z} \) represents the translation index. The normalization factor \(2^{l/2}\) ensures that the generated functions preserve orthonormality in the Hilbert space \(L^2(\mathbb{R})\). In this manner, the family \( \{\psi_{l,k}\}_{l,k \in \mathbb{Z}} \), combined with the basis generated by \( \{\phi_{l_{0},k}\}_{k \in \mathbb{Z}} \) for a fixed level \( l_{0} \), constitutes an orthonormal basis of \(L^2(\mathbb{R}) \). This allows functions in this class to be represented as weighted sums of these wavelets. 

\begin{figure}[!hbt]	
	\centering
	\includegraphics[width=1\textwidth]{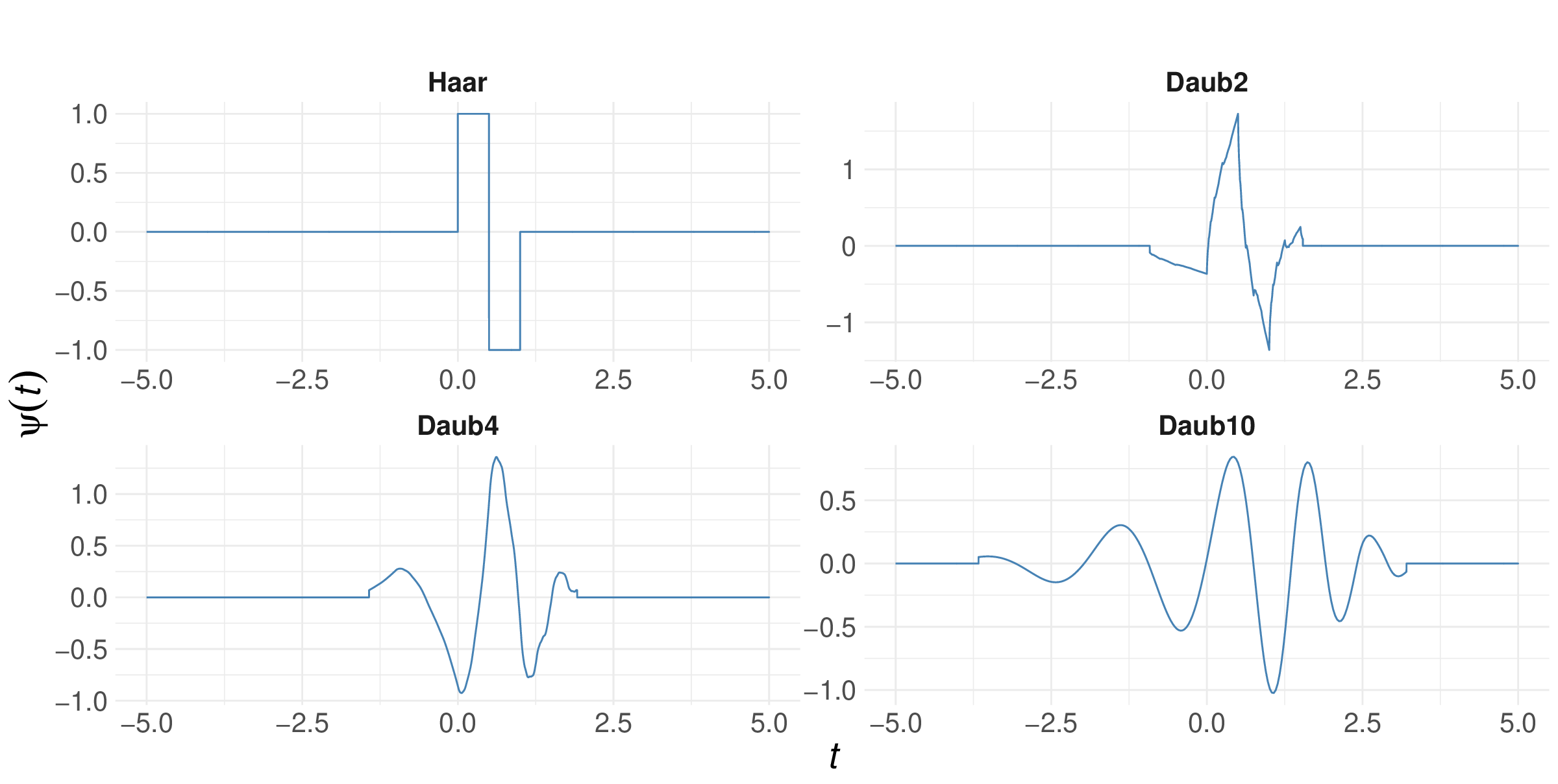}
	\caption{Daubechies wavelet functions with N = 1 (Haar wavelet), 2 (Daub2), 4 (Daub4) and 10 (Daub10) vanishing moments.}
	\label{fig:1.1}
\end{figure}

This construction is part of multiresolution analysis (MRA), a theoretical framework that organizes functions into various levels of detail or resolution. In this context, the scaling function \(\phi\) generates a sequence of approximation spaces \(V_l\) that capture the smooth or global characteristics of functions. Conversely, the wavelets \(\psi_{l,k}\) create the detail spaces \(W_l\), responsible for describing the fluctuations or irregularities that occur between successive resolution levels. Formally, this structure leads to the orthogonal decomposition: \(L^2(\mathbb{R}) =\displaystyle \bigoplus_{l \in \mathbb{Z}} W_l,\) with \(V_{l+1} = V_l \oplus W_l\), where the symbol \(\oplus\) denotes orthogonal direct sum. This formalism allows us to represent any function \(f \in L^2(\mathbb{R})\) as \(\displaystyle f(t) = \sum_{k} \alpha_{l_0, k} \, \phi_{l_0, k}(t) + \sum_{l = l_0}^{\infty} \sum_{k} \beta_{l,k} \, \psi_{l,k}(t)\), where the coefficients $\alpha_{l_0,k}$ represent the coarse approximation of the function at a base level $l_0$. The coefficients $\beta_{l,k}$ describe the progressive details of the function at finer levels. There are several wavelet functions; examples are shown in Figure~\ref{fig:1.1} for Daubechies wavelets with one (Haar or Daub1), two (Daub2), four (Daub4), and ten (Daub10) vanishing moments. For a complete and rigorous mathematical description, refer to several classic works, such as those by \citet{mallat:1989a,mallat:1989b}, \citet{daubechies:1992}, \citet{meyer:1992}, \citet{vidakovic:1999}, and \citet{mehra:2018}.
\subsection{Statistical Model}\label{sec:2.2}
Let \(s_j \in D \subset \mathbb{R}^d\) represent the spatial location of the \(j\)-th function, where \( j=1, \ldots, m \) and typically \(d=2\) in most applications. Consider \(t \in T = [a, b] \subset \mathbb{R}\) as the common continuum domain for all functions, which is discretized into \( n = 2^L \) equally spaced points, denoted as \(t = t_1, \ldots, t_n \). The functional observation corresponding to location \(s_j\) at time \(t\) is denoted by \(Y_{s_j}(t)\). We assume a non-parametric regression model to describe this configuration
\begin{align}\label{eq:npmodel}
    Y_{s_j}(t)=X_{s_j}(t)+\varepsilon_{s_j}(t).
\end{align}
The term \(\varepsilon_{s_j}(t)\) represents independent noise that follows a normal distribution, specifically \(\text{Normal}(0, \sigma_{\varepsilon}^2)\). This noise is independent across different spatial locations \(s_j\) and different time points \(t_i\) for \(i = 1, \ldots, n\). The latent function  \(X_{s_j}(\cdot)\), defined over the interval  \([a, b]\), belongs to the space \(L^2([a, b])\) and can be approximated by an expansion in a wavelet basis involving multiple levels of resolution and translation, indexed by \(l\) and \(k\), respectively
\begin{align}
    X_{s_j}(t)\approx\sum_{k=0}^{2^{l_{0}}-1}\alpha_{j,l_{0},k}\phi_{l_{0},k}(t)+\sum_{l=l_{0}}^{L-1}\sum_{k=0}^{2^{l}-1}\beta_{j,l,k}\psi_{l,k}(t),
\end{align}
where \(\{\phi_{l_0,k}\}_{k=0}^{2^{l_0}-1}\) are scaling functions at the coarsest level \(l_0\) and \(\{\psi_{l,k}\}\) are wavelet functions at levels \(l = l_0, \ldots, L-1\). We collect the scaling coefficients at level \(l_0\) into \(\boldsymbol\alpha_j = (\alpha_{j,l_0,0}, \ldots, \alpha_{j,l_0,2^{l_0}-1})^\top \in \mathbb{R}^{2^{l_0}}\). For each finer level \(l \ge l_0\), we write \(\boldsymbol\beta_{j,l} = \left(\beta_{j,l,0}, \ldots, \beta_{j,l,2^l-1}\right)^\top \in \mathbb{R}^{2^l}\). Collectively, we refer to \(\boldsymbol\theta_j = \left(\boldsymbol\alpha_j, \boldsymbol\beta_{j,l_0}, \boldsymbol\beta_{j,l_0+1}, \ldots, \boldsymbol\beta_{j,L-1}\right)\) as the full wavelet coefficient vector for \(X_{s_j}(t)\).

Let \(\mathbf{W}\) be the orthonormal matrix of the discrete wavelet transform (DWT) of dimension \(n\times n\), corresponding to the decomposition levels \(l_0, l_0 + 1, \ldots, L - 1\). By applying this matrix to the vectorized observations at location $s_j$, denoted \(\mathbf{Y}_{s_j} = \left(Y_{s_j}(t_1), \ldots, Y_{s_j}(t_n)\right)^{\top}\), we obtain the vector of empirical wavelet coefficients, given by \(\mathbf{d}_j = \mathbf{W}^\top \mathbf{Y}_{s_j}\). In matrix form, the measurement model can be expressed as follows:
\begin{align}
\mathbf{Y}_{s_j} \;=\;\mathbf{W}\,\boldsymbol\theta_j \;+\;\bm{\varepsilon}_j,
\quad
\bm{\varepsilon}_j\sim\text{Normal}\bigl(\mathbf{0},\,\sigma_{\varepsilon}^2\mathbf{I}_n\bigr),
\end{align}
and since $\mathbf{W}$ is orthonormal, we have
\begin{align}
   \mathbf{d}_j
\;=\;\mathbf{W}^\top\mathbf{Y}_{s_j}
\;=\;\boldsymbol\theta_j \;+\;\boldsymbol\varepsilon_j^*, 
\quad
\boldsymbol\varepsilon_j^*=\mathbf{W}^\top\boldsymbol\varepsilon_j\sim\text{Normal}\bigl(\mathbf{0},\,\sigma_{\varepsilon}^2\mathbf{I}_n\bigr). 
\end{align}
Thus, in the wavelet domain, each location \(s_j\) produces observed coefficients \(\mathbf{d}_j \in \mathbb{R}^n\) that consist of the true coefficient vector \(\boldsymbol\theta_j\) plus independent noise \(\boldsymbol\varepsilon_j^*\).
\subsection{Bayesian Hierarchical Model and the Regularized Horseshoe Prior}\label{sec:rhs}
In \citet{song:2019}, a specific spatial process is assigned to each pair of indices $(l, k)$, applied to a vector of wavelet coefficients. In this scheme, $\boldsymbol{\Sigma}_{l,k} \in \mathbb{R}^{m \times m}$ represents a particular covariance matrix associated with the level $l$ and the translation $k$. Although this approach provides remarkable flexibility in modeling spatial dependence, it involves estimating $\sum_{l=l_0}^{L-1} 2^l$ distinct covariance matrices, which incurs a high computational cost and increases the risk of identifiability issues for the spatial parameters. As an alternative, we propose a more parsimonious approach in which all coefficients corresponding to the same level $l$ share a single spatial process.

\textbf{The first stage (data):} For each level \( l = l_0, \dots, L-1 \), and each translation \( k = 0, \dots, 2^l - 1 \), we define the vector of empirical coefficients that correspond to all spatial locations as \(\mathbf{d}_{\cdot lk} = (d_{1lk}, \dots, d_{mlk})^\top\), which follows:
\begin{align*}
\mathbf{d}_{\cdot, l, k} \mid \boldsymbol{\theta}_{\cdot, l, k}, \sigma^2_{\varepsilon} \sim \text{Normal}_m(\boldsymbol{\theta}_{\cdot, l, k}, \sigma^2_{\varepsilon} I_m).
\end{align*}

\textbf{The second stage (spatial process and sparsity):} In wavelet-based modeling, it is well known that the coefficients capture various features of the underlying function at different resolution levels. Coarse levels reflect the global, smooth characteristics of the functions, where most coefficients are expected to be non-zero. In contrast, fine-resolution levels represent highly localized fluctuations and are typically sparse. These levels often contain mostly noise, with only a few non-zero coefficients corresponding to true sharp features. Traditionally, sparsity in Bayesian wavelet models has been induced using discrete spike-and-slab priors, which are mixtures of a point mass at zero and a continuous distribution. However, exploring the resulting highly multimodal posterior space poses significant computational challenges for gradient-based sampling algorithms, such as HMC and NUTS, as implemented in \texttt{Stan}. To ensure both theoretical identifiability and computational feasibility, we propose a level-dependent continuous shrinkage strategy that utilizes the RHS prior \citep{piironen:2017}.

A crucial modeling decision is to divide the resolution levels into two sets: coarse levels, denoted by \(l \le L_c\) (modeled as strictly dense), and fine levels, denoted by \(l > L_c\) (undergoing shrinkage). To determine the threshold \(L_c\) separating these two sets, we use the primary resolution-level formulation proposed by \citet{angelini:2004}. Based on the asymptotic considerations established by \citet{hardle:1998}, the number of coarse levels is explicitly defined as \(L_c = \lfloor \log_2(\log n) \rfloor + 1\). This threshold effectively identifies the initial levels at which the sparsity assumption no longer holds, thereby safeguarding the fundamental shape and primary energy of the curves from excessive regularization. Furthermore, integrating continuous shrinkage with spatial dependence requires a modeling choice that the coupling itself renders unavoidable. If we model the spatial process using a standard full covariance matrix \(\boldsymbol{\Sigma}_l = \sigma_l^2 \boldsymbol{R}_l\), the total variance of a fine-level coefficient becomes the product \(\sigma_l^2 \times \tau_l^2 \tilde{\lambda}_{l,k}^2\). This induces a non-identifiability issue, as the model cannot differentiate between a large spatial variance ($\sigma_l^2$) and a strong local curve feature captured by the shrinkage penalty ($\tilde{\lambda}_{l,k}^2$), since both scale the same coefficients. This confounding is distinct from the decay-versus-variance identifiability concerns associated with assigning a separate spatial process to every coefficient \citep{song:2019}; it arises specifically from the interaction between continuous shrinkage and spatial variance, an interaction that, to our knowledge, has not been made explicit in the wavelet RHS setting. To resolve it, we restrict the Matérn function to a spatial correlation matrix $\boldsymbol{R}_l  \in \mathbb{R}^{m \times m}$, fixing its marginal variance to $1$, so that the level-wise shrinkage scale alone governs the amplitude of the coefficients. 

The spatial dependence among wavelet coefficients is modeled using an isotropic Matérn correlation function. For each resolution level \( l \), the correlation matrix \(\boldsymbol{R}_l\) is defined as  \(\boldsymbol{R}_l[j, j'] =\rho(h_{j, j'}; \varphi_l, \nu)\), where \( h_{j,j'} = ||s_j - s_{j'}|| \) represents the Euclidean distance between the spatial locations \( s_j \) and \( s_{j'} \). The Matérn family of correlation functions \citep{diggle:2007} is given by:
\begin{align*}
\rho(h; \varphi_l, \nu) = \frac{1}{2^{\nu - 1} \Gamma(\nu)} (h/\varphi_l )^{\nu} K_{\nu}(h/\varphi_l).
\end{align*}
In this formulation, the parameter \(\varphi_l > 0\) represents the spatial decay (or range) parameter, while \(\nu > 0\) controls the smoothness of the field. The value of \(\nu\) determines the degree of mean-square differentiability of the trajectories of the underlying Gaussian process. By setting \(\nu = 1.5\), we ensure that the trajectories are once differentiable in \(L^2\). This specific choice also yields a closed-form expression, enhancing the computational efficiency of the inferential procedure \citep{stein:1999,diggle:2002,gneiting:2010}. Additionally, \(\Gamma(\cdot) \) is the gamma function, and \( K_{\nu}(\cdot) \) the modified Bessel function of the second kind of order \(\nu\). 

For coarse approximation levels (where \( l \leq L_c \)), the spatial coefficient vectors for each translation \( k \) at the same resolution level are modeled straightforwardly as
\begin{align} \label{eq:coarse}
    \theta_{\cdot, l,k} \mid \sigma^{2}_{l}, \boldsymbol{R}_l \sim \text{Normal}_m(0, \sigma_{l}^2 \boldsymbol{R}_l),
\end{align}
where $\sigma_{l} > 0$ is a level-specific scale parameter that governs the overall amplitude of the dense coarse signal. For the sparse fine detail levels ($l > L_c$), the coefficient vectors are modeled using the multivariate extension of the RHS
\begin{align}\label{eq:fine}
\theta_{\cdot, l, k} \mid \lambda_{l, k}, \tau_l, c_l, \boldsymbol{R}_l \sim \text{Normal}_m(0, \tau_l^2 \tilde{\lambda}_{l, k}^2 \boldsymbol{R}_l).
\end{align}
In this context, $\tau_l$ serves as the global shrinkage parameter for level $l$, which pulls all coefficients toward zero. On the other hand, $\lambda_{l, k}$ functions as the local shrinkage parameter, enabling individual translation coefficients to move away from the global shrinkage when the data strongly indicate the presence of a true local feature. The modified local penalty, denoted as $\tilde{\lambda}_{l, k}^2$, is defined as \(\displaystyle\tilde{\lambda}_{l, k}^2 = {\left(c_l^2 \lambda_{l, k}^2 \right)}/{\left(c_l^2 + \tau_l^2 \lambda_{l, k}^2\right) }\). This formulation ensures that for small signals, where \(\tau_l^2 \lambda_{l,k}^2 \ll c_l^2\), the prior functions similarly to the standard horseshoe, effectively reducing the noise to zero. In contrast, for large true signals, where \(\tau_l^2 \lambda_{l,k}^2 \gg c_l^2\), the prior variance approaches \(c_l^2 \boldsymbol{R}_l\), which helps prevent excessive smoothing of sharp features in the curve. The parameter \(c_l\) serves as the slab width. To complete the hierarchical specification, we assign independent Half-Cauchy marginal priors to both the local and global shrinkage parameters, and an Inverse-Gamma prior to the slab scale. The explicit theoretical structure is provided by
\begin{equation*}
\begin{gathered}
\lambda_{l,k} \sim \text{Half-Cauchy}(0, 1), \quad 
\tau_l \mid \tau_{0,\,l} \sim \text{Half-Cauchy}(0, \tau_{0,l}), \\[0.1cm]
c_l^2 \sim \text{Inv-Gamma}\left(\nu_c / 2, \nu_c\, s_c^2 / 2\right),
\end{gathered}
\end{equation*}
where the slab hyperparameters are fixed at $s_{c} = 2.0$ for the scale and $\nu_{c} = 4.0$ for the degrees of freedom. While this marginal formulation clearly defines the shrinkage behavior, directly sampling from heavy-tailed Half-Cauchy distributions can lead to problematic funnel shapes in HMC. To facilitate effective exploration of the posterior space, these priors are applied using a strictly Non-Centered Parameterization (NCP), which is explained in detail in Section \ref{sec:Computational Setup and Bayesian Inference}. Finally, instead of treating the global shrinkage scale \(\tau_{0,\,l}\) as completely unknown, we calibrate it using prior information about the expected sparsity. Following the approach of \citet{piironen:2017}, we inform the scale parameter \(\tau_{0,\,l}\) analytically based on the effective number of non-zero parameters \(p_{0,\,l}\) that we expect a priori at level \(l\). Specifically, \(\tau_{0,\,l}\) is defined as \(\displaystyle \frac{p_{0,\,l}}{D_l - p_{0,\,l}}\frac{\sigma_{\varepsilon}}{\sqrt{m}}\), where \(D_l = 2^l\) represents the total number of translation coefficients at that level, $m$ is the number of spatial locations, and $\sigma_{\varepsilon}$ is the observation noise standard deviation. As the resolution level \(l\) increases, \(p_{0,\,l}\) is designed to reflect an increasingly smaller fraction of the total \(D_l\) coefficients. This approach naturally incorporates the multiresolution decay of energy into the data-driven regularization scheme.

\textbf{In the third stage}, the hierarchical model is completed by assigning weakly informative prior distributions to the hyperparameters that control spatial dependence and amplitude scales
\begin{equation*}
\begin{gathered}
\varphi_l \sim \text{Lognormal}(\log(0.35 \cdot \text{d}_{max}), 1), \quad 
\sigma_{l} \sim \text{Half-Student-}t(4, 0, 1), \\[0.1cm]
\sigma_{\varepsilon} \sim \text{Half-Student-}t(4, 0,1).
\end{gathered}
\end{equation*}
Here, \( d_{\text{max}} = \max_{j,j'} \|s_j - s_{j'}\| \) represents the maximum observed Euclidean distance between any pair of spatial locations in the domain. The lognormal prior for the spatial decay parameter \( \varphi_l \) is intentionally centered around \( \log(0.35 \cdot d_{\text{max}}) \). This heuristic formulation ensures that the effective spatial range spans a plausible fraction of the domain, thereby preventing the Matérn correlation structure from either degenerating into pure spatial independence (when the range approaches zero) or becoming indistinguishable from a global flat field (when the range approaches infinity). For the scale parameters governing the dense coarse signal ($\sigma_l$) and the observational noise ($\sigma_{\varepsilon}$), Half-Student-$t$ distributions serve as weakly informative priors with heavier tails than the Gaussian. This property lets the data, rather than the prior, drive the estimation of large-variance components while, for a moderate number of degrees of freedom, retaining finite variance that maintains well-behaved posterior geometry under gradient-based sampling. This choice also avoids the artificial concentration away from zero that Inverse-Gamma priors can impose on scale parameters in hierarchical models.
\subsection{Spatial Prediction at Unobserved Locations}\label{sec:prediction}
A key objective of the proposed framework is to predict complete functional trajectories at locations with no observed data. The estimation process is performed using the NUTS algorithm in \texttt{Stan}, as detailed in Section~\ref{sec:Computational Setup and Bayesian Inference}. However, prediction requires no additional posterior simulation. This subsection demonstrates that once the joint posterior is obtained, the predictive distribution can be expressed in a closed conditional form. Consequently, prediction simplifies to a direct sampling step, which we implement within the same Stan run. This step can also be replicated as a post-processing task in any environment.

Let \(\mathcal{S}_o = \{s_1, \ldots, s_m\}\) represent the observed locations, and let \(\mathcal{S}_{\ast} = \{s^{\ast}_1, \ldots, s^{\ast}_q\}\) denote a set of \(q\) new, unsampled locations where we aim to make predictions. Our prediction process operates directly within the multiresolution domain rather than in a continuous functional space. For each resolution level \(l\) and translation \(k\), we augment the coefficient vector \(\bm{\theta}_{\cdot,l,k} \in \mathbb{R}^{m}\) with its unobserved counterpart \(\bm{\theta}^{\ast}_{\cdot,l,k} \in \mathbb{R}^{q}\). A key observation is that both the coarse-level prior (as shown in Equation \eqref{eq:coarse}) and the fine-level Regularized Horseshoe prior (as seen in Equation \eqref{eq:fine}) are zero-mean multivariate normal distributions. The covariance of these distributions factors into a scalar scale times a level-specific Matérn correlation matrix. We define this scalar as follows:
\begin{equation}
v_{l,k} =
\begin{cases}
\sigma_l^{2}, & l \leq L_c \quad \text{(dense coarse levels)},\\[3pt]
\tau_l^{2}\,\tilde{\lambda}_{l,\,k}^{2}, & l > L_c \quad \text{(sparse fine levels)}.
\end{cases}
\label{eq:vlk}
\end{equation}
In both cases, we can express the relationship as \(\bm{\theta}_{\cdot,l,k} \mid \Xi \sim \mathrm{Normal}_m(\mathbf{0}, v_{l,k}\mathbf{R}_l)\), where \(\Xi\) represents all hyperparameters of the hierarchy discussed in Section~\ref{sec:rhs}. This unified expression arises from restricting the Matérn function to a correlation matrix. The matrix \(\mathbf{R}_l\) does not contribute any variance itself; thus, the overall scale of the process whether derived from the coarse amplitude \(\sigma_l\) or from the shrinkage profile \(\tau_l\tilde{\lambda}_{l,k}\) is encapsulated in the single scalar \(v_{l,k}\). This separation simplifies the prediction process, making it clearer and more computationally efficient. During the implementation phase, \(\sqrt{v_{l,k}}\) is the per-coefficient scale that was already established when constructing the latent field during estimation. As a result, we do not need to recompute any quantities specifically for prediction. Extending the Matérn correlation function to all pairwise distances among \(\mathcal{S}_o \cup \mathcal{S}_{\ast}\) yields the augmented matrix \(\mathbf{R}^{+}_{l} \in \mathbb{R}^{(m+q)\times(m+q)}\), partitioned in blocks as
\begin{equation}
\mathbf{R}^{+}_{l} =
\begin{pmatrix}
\mathbf{R}_{oo} & \mathbf{R}_{o\ast} \\[2pt]
\mathbf{R}_{\ast o} & \mathbf{R}_{\ast\ast}
\end{pmatrix},
\qquad
\mathbf{R}_{oo} \in \mathbb{R}^{m \times m}, \quad
\mathbf{R}_{\ast\ast} \in \mathbb{R}^{q \times q}, \quad
\mathbf{R}_{\ast o} = \mathbf{R}_{o\ast}^{\top},
\label{eq:Rplus}
\end{equation}
where $\mathbf{R}_{oo}$ collects the correlations among observed sites, $\mathbf{R}_{\ast\ast}$ those among the new sites, and $\mathbf{R}_{\ast o}$ the cross-correlations between the two sets. Every block is evaluated from the same level-specific decay $\varphi_l$, so that no additional parameter is introduced by the prediction step. Under the augmented hierarchy, the joint prior of the stacked vector is
\begin{equation}
\begin{pmatrix} \bm{\theta}_{\cdot,l,k} \\[2pt]
\bm{\theta}^{\ast}_{\cdot,l,k} \end{pmatrix} \;\Big|\; \Xi \;\sim\;
\mathrm{Normal}_{m+q}\!\left(\mathbf{0},\; v_{l,k}\,\mathbf{R}^{+}_{l}\right).
\label{eq:jointprior}
\end{equation}
Applying standard multivariate normal conditioning to~\eqref{eq:jointprior}
gives, for every pair $(l,k)$,
\begin{equation}
\bm{\theta}^{\ast}_{\cdot,l,k} \mid \bm{\theta}_{\cdot,l,k}, \Xi \;\sim\;
\mathrm{Normal}_q\big(\mathbf{m}_{l,k},\, \mathbf{V}_{l,k}\big),
\label{eq:condpred}
\end{equation}
with conditional mean and covariance
\begin{equation}
\mathbf{m}_{l,k} = \mathbf{R}_{\ast o}\,\mathbf{R}_{oo}^{-1}\,
\bm{\theta}_{\cdot,l,k},
\qquad
\mathbf{V}_{l,k} = v_{l,k}\big(\mathbf{R}_{\ast\ast} -
\mathbf{R}_{\ast o}\,\mathbf{R}_{oo}^{-1}\,\mathbf{R}_{o\ast}\big).
\label{eq:condmoments}
\end{equation}
Note that the scale \(v_{l,k}\) cancels in the mean but survives in the covariance. This has three consequences. The mean is a simple-kriging predictor in coefficient space, with interpolation weights
\(\mathbf{R}_{\ast o}\mathbf{R}_{oo}^{-1}\) that depend only on \(\varphi_l\) and are shared by all translations within the level. The covariance, by retaining \(v_{l,k}\), inherits the model's regularization: a coefficient shrunk toward zero at the observed sites predicts near zero at the new sites, whereas one that escaped shrinkage remains diffuse, so the adaptive sparsity learned in estimation propagates coherently to prediction. Finally, since \(\mathbf{R}_{oo}\) depends only on \(\varphi_{l}\), one Cholesky factorization per level serves all \(2^{l}\) translations, \(\mathcal{O}(L)\) factorizations per draw, versus \(\mathcal{O}(n)\) under a coefficient-specific construction.

\paragraph{Posterior predictive distribution.} Let \(\mathbf{Y}_{1:m}\) denote the observed curves and \(\bm{\theta}\), \(\bm{\theta}^{\ast}\) the coefficients at the observed and new locations. Since a latent curve is recovered deterministically as $\mathbf{X}^{\ast} = \mathbf{W}\bm{\theta}^{\ast}$, integrating the conditional law~\eqref{eq:condpred} over the joint posterior gives the posterior predictive distribution of the latent curve,
\begin{equation}
p\big(\mathbf{X}^{\ast} \mid \mathbf{Y}_{1:m}\big) =
\int p\big(\mathbf{X}^{\ast} \mid \bm{\theta}^{\ast}\big)\,
p\big(\bm{\theta}^{\ast} \mid \bm{\theta}, \Xi\big)\,
\pi\big(\bm{\theta}, \Xi \mid \mathbf{Y}_{1:m}\big)\,
d\bm{\theta}^{\ast}\, d\bm{\theta}\, d\Xi .
\label{eq:postpred}
\end{equation}
The latent curve \(\mathbf{X}^{\ast}\) represents the noise-free process at the new site and serves as the point predictor throughout the analysis. A new observation, \(\mathbf{Y}^{\ast}\), reflects the value recorded by a sensor and includes measurement error, as described in Equation~\eqref{eq:npmodel}. Specifically, \(\mathbf{Y}^{\ast} \mid \bm{\theta}^{\ast}, \sigma_{\varepsilon} \sim \mathrm{Normal}_n(\mathbf{W}\bm{\theta}^{\ast}, \sigma_{\varepsilon}^{2}\mathbf{I}_n)\), which results in wider prediction intervals. Which target is used for coverage depends on what is knowable. In the simulation study, the noise-free curve \(Z_j\) is known, so coverage is assessed for \(\mathbf{X}^{\ast}\) against \(Z_j\); in the application, the latent process is never observed, so coverage is assessed for \(\mathbf{Y}^{\ast}\) against the held-out measurements. In each case, the interval and target share the same noise structure, which the alternative pairing would violate.

The integral presented in Equation \eqref{eq:postpred} is difficult to compute directly. However, it can be approximated by using composition sampling from the closed-form Gaussian shown in Equation \eqref{eq:condpred}, without requiring additional posterior simulations. In our implementation, this process is carried out within the generated-quantities block, but it can also be performed as a post-processing step on saved samples. The procedure is summarized in Algorithm \ref{alg:pred}.
\begin{algorithm}[!ht]
\caption{Posterior predictive sampling of curves at unobserved locations}
\label{alg:pred}
\begin{algorithmic}[1]
\Require Posterior draws $\{(\bm{\theta}^{(g)}, \Xi^{(g)})\}_{g=1}^{G}$;
locations $\mathcal{S}_o$, $\mathcal{S}_{\ast}$; basis $\mathbf{W}$; target
($\mathbf{X}^{\ast}$ or $\mathbf{Y}^{\ast}$)
\For{$g = 1, \ldots, G$}
  \For{$l = l_0, \ldots, L-1$}
    \State Partition $\mathbf{R}^{+(g)}_{l}$ from $\varphi_l^{(g)}$
           as in~\eqref{eq:Rplus}; factorize $\mathbf{R}_{oo} =
           \mathbf{L}^{(g)}_{l}(\mathbf{L}^{(g)}_{l})^{\top}$
           \Comment{once per level}
    \State $\mathbf{K}^{(g)}_{l} \gets (\mathbf{L}^{(g)}_{l})^{-1}
           \mathbf{R}_{o\ast}$; \;
           $\mathbf{S}^{(g)}_{l} \gets \mathbf{R}_{\ast\ast} -
           (\mathbf{K}^{(g)}_{l})^{\top}\mathbf{K}^{(g)}_{l}$
    \For{$k = 0, \ldots, 2^{l}-1$}
      \State $v_{l,k}^{(g)} \gets$ Eq.~\eqref{eq:vlk}; \;
             $\mathbf{m}_{l,k}^{(g)} \gets (\mathbf{K}^{(g)}_{l})^{\top}
             (\mathbf{L}^{(g)}_{l})^{-1}\bm{\theta}^{(g)}_{\cdot,l,k}$
      \State Draw $\bm{\theta}^{\ast(g)}_{\cdot,l,k} \sim
             \mathrm{Normal}_q(\mathbf{m}_{l,k}^{(g)},
             v_{l,k}^{(g)}\mathbf{S}^{(g)}_{l})$
    \EndFor
  \EndFor
  \State $\mathbf{X}^{\ast(g)} \gets \mathbf{W}\bm{\theta}^{\ast(g)}$; \;
         if target is $\mathbf{Y}^{\ast}$, add
         $\bm{\varepsilon}^{(g)} \sim \mathrm{Normal}_n(\mathbf{0},
         \sigma_{\varepsilon}^{2(g)}\mathbf{I}_n)$
  \EndFor
\State Summarize the draws pointwise (predictive mean)
\end{algorithmic}
\end{algorithm}

Two additional notes complete the description. The algorithm is applied to the padded coefficients under the symmetric padding method described in Section~\ref{sec:Computational Setup and Bayesian Inference}. The mirrored regions are trimmed only after the inverse transform, which helps maintain the structure of the dyadic wavelet tree. To ensure numerical stability, jitter is added to the diagonals of $\mathbf{R}_{oo}$ and $\mathbf{S}^{(g)}_{l}$, and the latter is symmetrized. We use the algorithm in Section~\ref{sec:sim3} and in the application discussed in Section~\ref{sec:realdata}.
\section{Simulation Studies}\label{sec:3}
To validate the proposed hierarchical spatial wavelet model, we conducted a comprehensive simulation study. Using synthetic data provides complete control over the data-generating process, enabling evaluation of the model's performance along three complementary dimensions. The first objective is to assess the signal-recovery capability of the spatial Regularized Horseshoe prior across varying levels of difficulty. For this purpose, we simulate functional datasets with strictly controlled noise variances, establishing distinct high-, moderate-, and low-SNR regimes (Section~\ref{sec:sim1}). The second objective is to quantify the empirical advantage of borrowing spatial strength. We evaluate this by comparing our spatial formulation against an Independent Regularized Horseshoe (IRHS) baseline (Section~\ref{sec:sim2}), which retains the same shrinkage prior but treats the $m$ functional curves as strictly independent, thereby neglecting the underlying spatial topology. The third objective is to evaluate predictive performance at unobserved locations (Section~\ref{sec:sim3}). Here, we benchmark our approach against two reference models: the classical OKFD method \citep{giraldo:2011} and the hierarchical Bayesian wavelet model of \citet{song:2019}. To assess robustness against model misspecification, the true latent spatially correlated coefficients are generated from an Exponential covariance function ($\nu = 0.5$), while inference is performed under a smoother Matérn~3/2 structure ($\nu = 1.5$). This design evaluates whether the model can recover the true functional parameters even when the assumed spatial smoothness differs from the data-generating mechanism.

 \subsection{Computational Setup and Bayesian Inference} \label{sec:Computational Setup and Bayesian Inference}
On the technical side, we implemented the simulation environment and data preprocessing in \texttt{R} \citep{R2026} using the \texttt{wavethresh} package. A common methodological challenge when applying the DWT to finite signals is boundary effects, which have been extensively documented in the wavelet literature \citep[see][Chapter 7]{Mallat:2008}. The standard periodic boundary condition (\texttt{bc="periodic"}) connects the end of the signal back to its beginning. However, if the values at these two endpoints differ significantly, this periodic wrapping creates an artificial discontinuity (a sharp jump). The DWT interprets this jump as high-frequency energy, resulting in spurious detail coefficients that can severely distort the Regularized Horseshoe shrinkage process at the boundaries. To mitigate this issue while strictly preserving the dyadic length (\(2^L\)) required by the hierarchical wavelet tree, we implemented a dynamic symmetric padding strategy. The original functional curves of length \(n\) were mirrored at both ends by prepending the reversed first half and appending the reversed second half-effectively creating a continuous, artifact-free extended sequence of length \(2n\). The DWT was then applied to these padded curves using Daubechies Extremal Phase wavelets with four vanishing moments (\texttt{family="DaubExPhase", filter.number=4}). After performing Bayesian estimation and the Inverse DWT, the padded regions were symmetrically trimmed, perfectly recovering the artifact-free signal in its original domain.

The statistical inference engine used in this study was \texttt{Stan} \citep{stan2025}, which employs the NUTS algorithm \citep{Hoffman2014}. This algorithm is a highly efficient Markov Chain Monte Carlo (MCMC) method. Given the extreme dimensionality of the latent wavelet space and the heavy-tailed nature of the Half-Cauchy distributions within the Regularized Horseshoe prior, we opted for a strict NCP for both the spatial Gaussian fields (\(\mathbf{z}\)) and the local shrinkage parameters. This architectural decision is crucial for decoupling the dependence between the location and scale hyperparameters. It helps to avoid the problematic "funnel" geometries that can obstruct convergence in complex hierarchical models \citep{Betancourt:2017}. We do not sample directly from the marginal Half-Cauchy distributions described in Section \ref{sec:rhs}. Instead, we implement the NCP by representing the shrinkage parameters as scale mixtures of Normal and Inverse-Gamma distributions. The structure of the latent variables, as coded for the MCMC, is defined as follows: For the local shrinkage parameters, we have \(\lambda_{l,k} = u_{1,\,l,k} \sqrt{u_{2,\,l,k}}\), where \(u_{1,\,l,k} \sim \text{Normal}(0, 1)\) and \(u_{2,\,l,k} \sim \text{Inv-Gamma}(0.5, 0.5)\). For the level-specific global shrinkage parameters, we define \(\tau_l = v_{1,\,l} \sqrt{v_{2,\,l}} \, \tau_{0,\,l}\), where \(v_{1,\,l} \sim \text{Normal}(0, 1)\) and \(v_{2,\,l} \sim \text{Inv-Gamma}(0.5, 0.5)\).

This structure prevents problematic funnel geometries in the joint posterior space, thereby improving the efficiency of the NUTS algorithm.

We assessed the frequentist properties of the Bayesian estimators in Simulation Studies 1 and 2 using a Monte Carlo (MC) scheme with 300 independent replicates. In Simulation Study 3, the computational cost associated with the \citet{song:2019} benchmark limited the Bayesian-versus-Bayesian comparison to the first 100 replicates, as described in Section~\ref{sec:sim3}. For each replicate and both Bayesian models, we approximated the posterior distribution using four parallel NUTS chains, each with 2,000 warm-up and 1,000 sampling iterations, yielding 4,000 posterior draws per parameter. To ensure robust exploration of the complex posterior geometry, the acceptance probability (\texttt{adapt\_delta}) was set to 0.99 and the maximum tree depth to 12. Convergence and sampling efficiency were evaluated using the rank-normalized potential scale reduction factor ($\hat{R} \le 1.01$) and the bulk and tail effective sample sizes (ESS) \citep{Vehtari2021}. The latter metric is particularly important due to the heavy-tailed Half-Cauchy components of the Regularized Horseshoe prior. All reported estimates met these diagnostic criteria.
 \subsection{Data Generation Process} \label{sec:3.2}
\begin{figure}[!hbt]	
	\centering
	\includegraphics[width=1\textwidth]{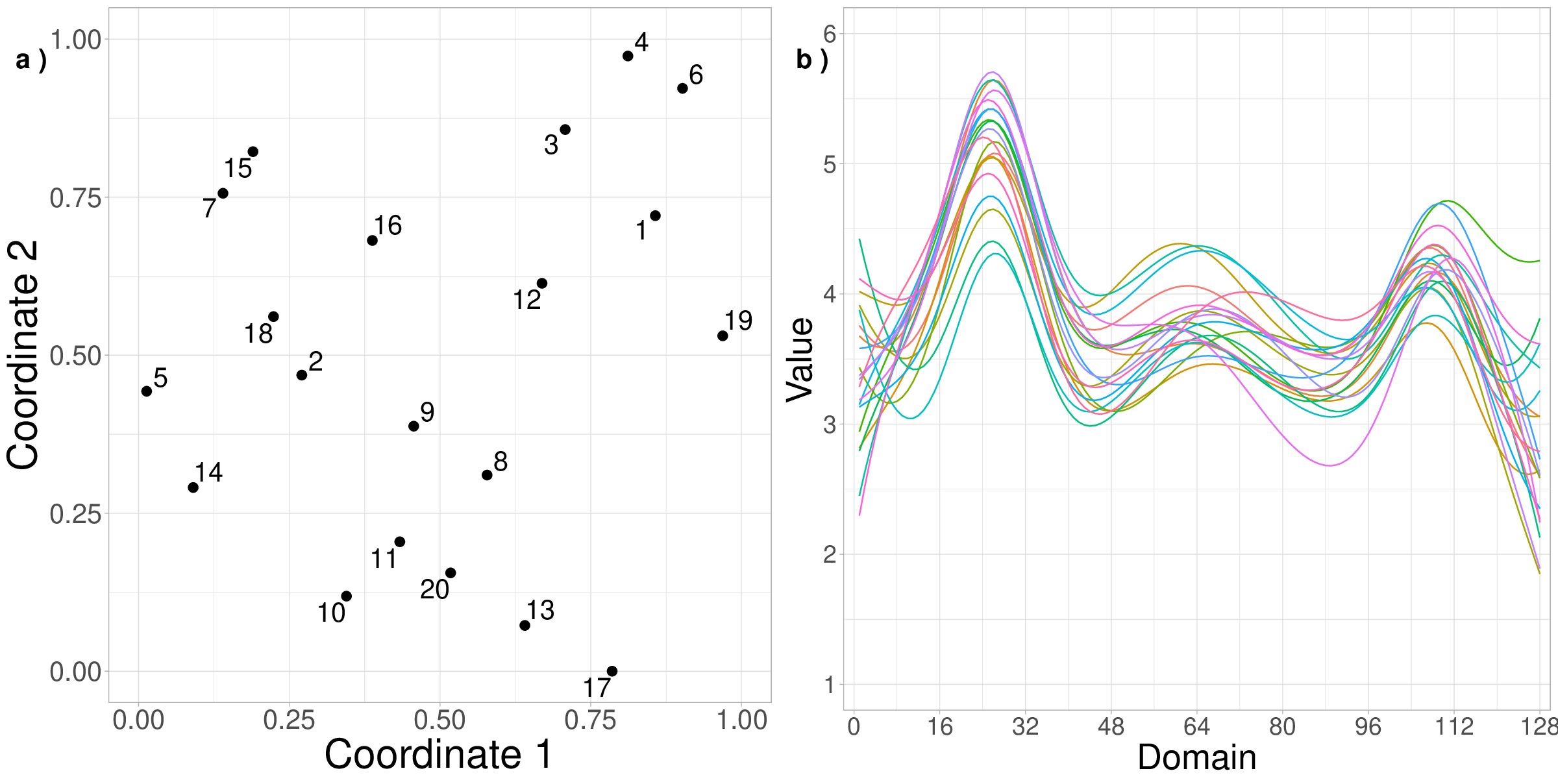}
	\caption{Graphs are from the simulation. Panel $(a)$ shows a grid of simulated locations. Panel $(b)$ presents the target curves without noise.}
	\label{fig:study1_data}
\end{figure}

To thoroughly evaluate the proposed hierarchical spatial wavelet model, we simulated functional data that replicate the complex, non-linear behaviors often observed in environmental and epidemiological time series. The simulation encompassed \( m = 20 \) spatial locations as shown in Panel (a) of Figure \ref{fig:study1_data}, with functional curves recorded at a regular discrete grid of \( n = 128 \) time points. The actual generated curves are depicted in Panel (b) of the same figure.

At each location \( s_j \) and time \( t \), the true underlying latent signal, denoted as \( Z_j(t) \), was constructed as the sum of a fixed deterministic trend and a spatially correlated random functional component. The deterministic trend, \( \mu(t) \), was designed as a trimodal function to produce distinct, sharp peaks (high-frequency features) alongside smooth valleys (low-frequency features). This base trend is defined by a global intercept and three Gaussian peaks, represented by the following equation
\begin{align*}
\mu(t) = 3.1 + &1.8\exp(-0.008(t-26)^2)\\\nonumber& + 0.8\exp(-0.003(t-64)^2)+ 1.0\exp(-0.008(t-109)^2). \end{align*}

To introduce functional variability across the spatial domain, we utilized a spatial basis expansion approach. We defined a set of \( K = 8 \) cubic B-spline basis functions (of order 4), denoted by \( {\Phi}(t) \). The latent curve for the \( j \)-th spatial location was then generated as follows
\begin{align*}
  Z_j(t) = \mu(t) + \sum_{k=1}^{K} a_{k,j} \Phi_k(t),  
\end{align*}
where the basis coefficients \( a_{kj} \) are spatially dependent. For each basis function \( k \), the vector of coefficients across all \( m \) locations, \( \bm{a}_k = (a_{k1}, \dots, a_{km})^\top \), was drawn from a multivariate normal distribution \( \bm{a}_k \sim \text{Normal}_m(\mathbf{0}, \boldsymbol{\Sigma}) \). The spatial covariance matrix \( \boldsymbol{\Sigma} \) was constructed using an exponential covariance function \(\Sigma_{j,j^{*}} = \sigma^2_s \exp\left( - d_{j,j^{*}}/\varphi_s \right)\), where $d_{j,j^{*}}$ is the Euclidean distance between locations $j$ and $j^{*}$. We fixed the marginal variance at $\sigma^2_s = 0.5$ and the spatial decay parameter at $\varphi_s = 0.4$, ensuring moderate spatial correlation.
\subsection{Simulation Study 1: Signal-to-Noise Ratio}\label{sec:sim1}
We created the synthetic datasets by adding independent Gaussian white noise to the true latent curves. This can be expressed mathematically as \(Y_j(t) = Z_j(t) + \varepsilon_j(t)\), \(\varepsilon_j(t) \sim \text{Normal}(0, \sigma^2_{\text{noise}})\). To systematically assess the effectiveness of the Regularized Horseshoe spatial prior in recovering the true signal, we designed three distinct simulation scenarios based on the SNR. For spatial functional data, the overall challenge of recovering the signal is best represented by the average curve-wise SNR across all spatial locations, which is defined as: $$\text{SNR} = \frac{1}{m} \sum_{j=1}^m \frac{\text{Var}(Z_j)}{\sigma^2_{\text{noise}}}.$$ This formulation allows us to quantify the relationship between the variance of the true signal and the variance of the noise, facilitating a comprehensive evaluation of the recovery task. By controlling the noise standard deviation (\(\sigma_{\text{noise}}\)), we designed the following Monte Carlo scenarios, each consisting of 300 independent replicates:
\begin{enumerate}
\item \textbf{Scenario 1 (High SNR / Low Noise)}: Setting \(\sigma_{\text{noise}} = 0.15\) results in an average SNR of approximately 14.41. This scenario tests the model's ability to maintain the true shape of the curves without over-smoothing or excessively flattening genuine signal peaks.
\item \textbf{Scenario 2 (Moderate SNR)}: With \(\sigma_{\text{noise}} = 0.30\), we obtain an average SNR of about 3.60. In this scenario, minor fluctuations in the function begin to blend with the noise, thereby challenging the model's thresholding capabilities.
\item \textbf{Scenario 3 (Low SNR / High Noise)}: Here, \(\sigma_{\text{noise}} = 0.60\) leads to an average SNR of approximately 0.90. This is the most challenging scenario, where the variance of the observational noise significantly overshadows that of the true signal. It serves as the ultimate stress test for assessing the mathematical advantage of leveraging information from spatial neighbors to recover heavily corrupted features.
\end{enumerate}

\begin{figure}[!hbt]	
	\centering
	\includegraphics[width=1\textwidth]{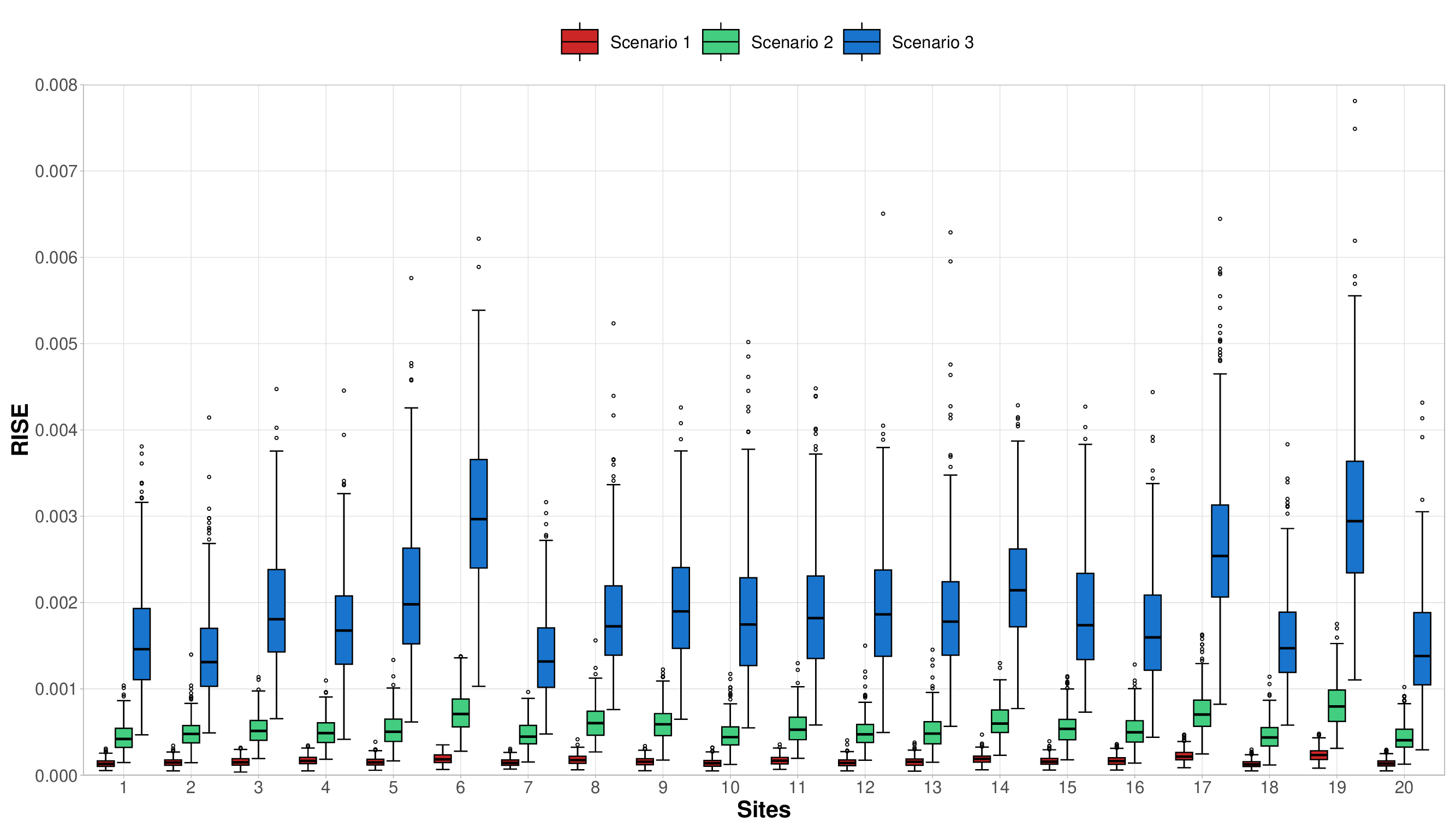}
	\caption{Distribution of the RISE across 20 locations from 300 MC simulations. Boxplots show the model's effectiveness at recovering target curves across three noise regimes: low, moderate, and high.}
	\label{fig:study1_boxplot}
\end{figure}

Figure \ref{fig:study1_boxplot} shows the distribution of the Relative Integrated Squared Error (RISE) obtained from 300 Monte Carlo simulations. This illustrates how effectively the spatial hierarchical formulation recovers the original target curves across three different SNRs. By standardizing the reconstruction error relative to the total energy of the true latent curves, the RISE yields a scale-free metric that enables clearer evaluation of the model's performance. For a detailed theoretical formulation of this discrepancy measure, refer to Appendix \ref{sec:app_metrics}.

In Scenario 1, which is characterized by a low-noise environment with a noise standard deviation of \(\sigma_{\text{noise}}=0.15\) and a high average SNR of approximately 14.41, the model demonstrates excellent accuracy. The RISE values are closely clustered around zero across all 20 locations. This outcome highlights the effectiveness of the spatial formulation of the RHS in maintaining the true shape of the curves by avoiding excessive smoothing and preventing the undesired flattening of the signal's genuine high-frequency peaks. Moreover, the local regularization parameter (\(\lambda_{l,k}\)) serves effectively to prevent the true detail coefficients from being globally shrunk to zero. In Scenario 2, with a moderate SNR of approximately 3.60 (\(\sigma_{\text{noise}} = 0.30\)), we observe a critical situation where smaller fluctuations in the function begin to merge with the background noise. This results in a natural yet carefully controlled increase in both the dispersion and the magnitude of the RISE values. However, the median relative error remains low, demonstrating the model's robustness in distinguishing true spatial variability from random fluctuations.

Finally, Scenario 3 presents the most severe stress test, evaluating a low-SNR environment in which the noise level rises to \(\sigma_{\text{noise}}=0.60\), resulting in an average SNR of approximately 0.90. In this situation, the observational noise magnitude significantly exceeds that of the underlying signal. As expected, both the overall magnitude and variance of the RISE reach their highest levels, with upper quartiles approaching 0.008 at certain locations. Nevertheless, maintaining a relative error of less than 1\% (RISE $\leq0.01$) under such extreme conditions underscores the necessity of the borrowing-of-spatial-strength mechanism. The model can still achieve effective functional reconstruction by employing continuous shrinkage guided by the Matérn correlation structure, thereby stabilizing heavily corrupted local features through information sharing across neighboring locations. 

\begin{figure}[!hbt]	
	\centering
	\includegraphics[width=1\textwidth]{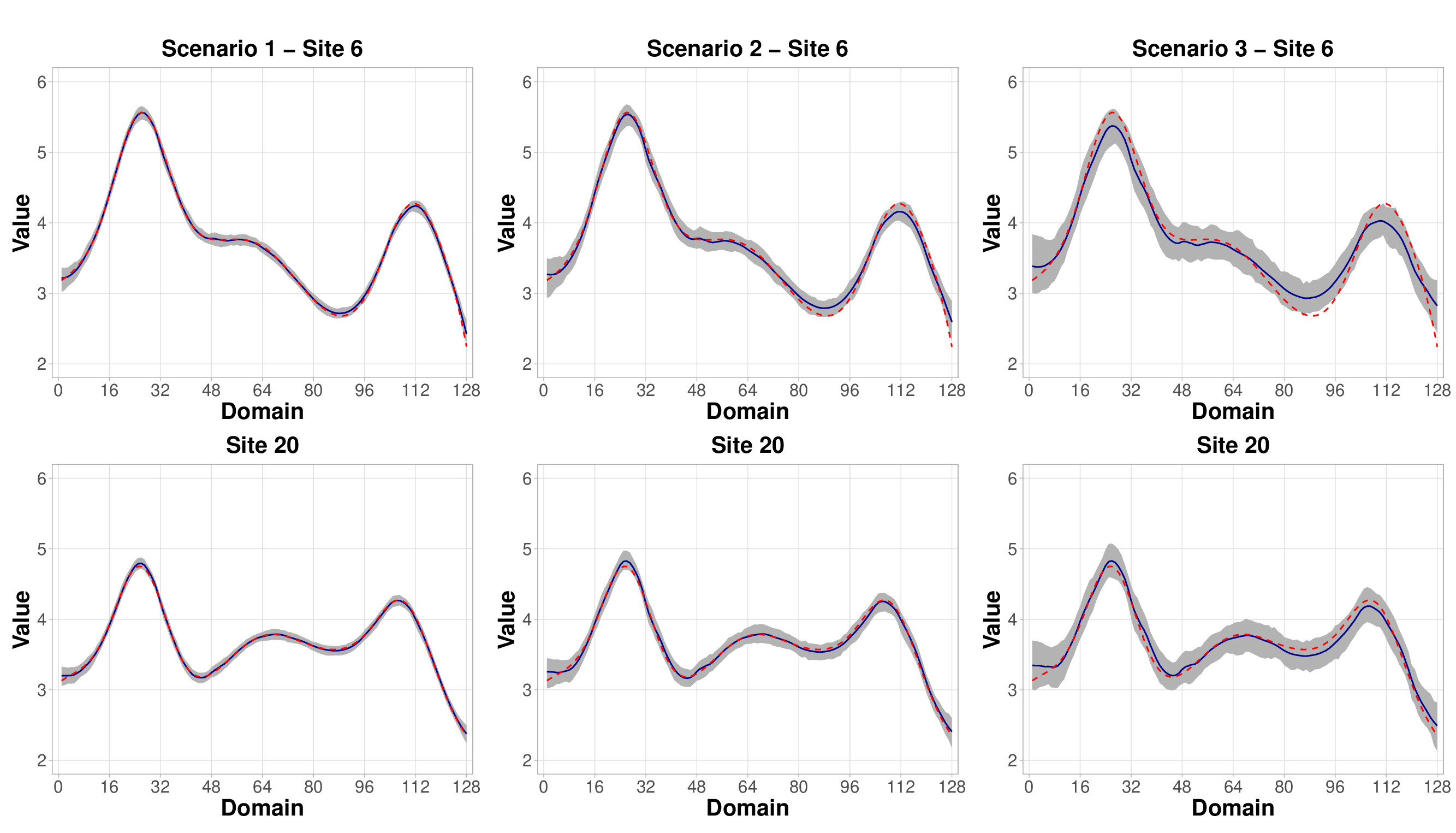}
	\caption{Functional reconstruction of the latent curves at Sites 6 and 20 across the three noise scenarios. Red dotted: true latent curve. Solid blue: point estimate (average of the 300 posterior-mean curves across MC replicates). Gray shaded: average of the 300 corresponding 95\% HPD intervals.}
	\label{fig:study1_boxplot2}
\end{figure}

Figure \ref{fig:study1_boxplot2} shows the proposed spatial model's ability to reconstruct the functional curves across three noise scenarios at two geographic locations: Sites 6 and 20. In this figure, the red dotted line represents the true underlying latent curve. To summarize the simulation study's overall performance, the solid blue line shows the point estimate, calculated as the average of the 300 smoothed posterior mean curves across Monte Carlo replicates. Additionally, the gray shaded area represents the average of the 300 corresponding 95\% Highest Posterior Density (HPD) credible intervals.

Selecting these two locations demonstrates the model's spatial robustness across various topological configurations. As shown in the spatial grid (Figure \ref{fig:study1_data}a), Site 6 is situated at the upper-right boundary of the domain, representing a peripheral area with fewer neighbors. In contrast, Site 20 is located in the lower central region. In Scenario 1 (left panels), which is characterized by a low-noise environment, the model captures the functional dynamics exceptionally well. The averaged blue posterior mean aligns closely with the true red curve, and the averaged 95\% HPD interval is remarkably narrow. This demonstrates that the RHS prior does not excessively penalize high-frequency coefficients when the signal is clear, thereby effectively preserving the amplitude of the three characteristic peaks of the original function. As the noise variance increases in Scenario 2, the average gray uncertainty band naturally widens to reflect reduced signal clarity. Despite this, the mean estimate remains remarkably stable and true to the original geometry, highlighting the hierarchical model's effectiveness as a robust thresholding mechanism against random fluctuations.

Scenario 3 (right panels) provides clear visual evidence of the model's effectiveness under extreme noise conditions ($\sigma_{\text{noise}}=0.60$). Even when observational variance significantly obscures the underlying signal, the averaged estimated curves show neither erratic oscillations (indicating overfitting) nor severe flattening (suggesting over-smoothing). The average 95\% HPD interval conservatively widens to fully encompass the true curve from end to end, ensuring accurate statistical coverage. Notably, even at Site 6, a location near a geometric boundary, the curve is reconstructed with good functional accuracy. This visually demonstrates the mechanics of spatial strength borrowing: the Matérn correlation structure compensates for poor local data quality by leveraging topological information from surrounding stations, enabling the RHS to reconstruct heavily corrupted features confidently.

\subsection{Simulation Study 2: Comparison of Spatial and Independent Models}\label{sec:sim2}
The goal of this second simulation study is to evaluate the statistical benefits of explicitly including spatial dependence when modeling correlated curves. To accomplish this, we compare our proposed hierarchical spatial model with a constrained version that assumes complete independence across all geographic locations. This baseline reference model, called the IRHS, has the same architecture, hierarchical structure, and continuous shrinkage priors as our proposed model. The key difference is that the Matérn spatial correlation matrix ($\boldsymbol{R}_l$) is completely removed and replaced with the identity matrix ($\boldsymbol{I}_m$). This change effectively eliminates any "borrowing of strength" between neighboring curves during inference.

To ensure a fair comparison, we keep all other structural components the same, including the NCP and the distinction between coarse and sparse levels. Thus, any improvement observed in predictive performance or parameter estimation of the proposed model can be directly attributed to its spatial framework. The comparative analysis focuses on the moderate noise regime, characterized by an average SNR of approximately 3.60. This aligns with the scenario where the noise standard deviation, \(\sigma_{\text{noise}}\), is set at 0.30, as detailed in Section \ref{sec:sim1}. This scenario is particularly insightful because it marks a critical threshold at which local, high-frequency features of the trimodal signal begin to merge with observational noise. Under these conditions, recovering the signal becomes highly challenging for independent estimators. Through 300 Monte Carlo simulations, we aim to demonstrate that the spatial model's ability to leverage neighborhood information yields more accurate estimates. Additionally, we quantify the deterioration in statistical quality, including increased reconstruction error and compromised credible-interval coverage, that occurs when spatial dependence is not accounted for.

\begin{figure}[!hbt]	
	\centering
	\includegraphics[width=1\textwidth]{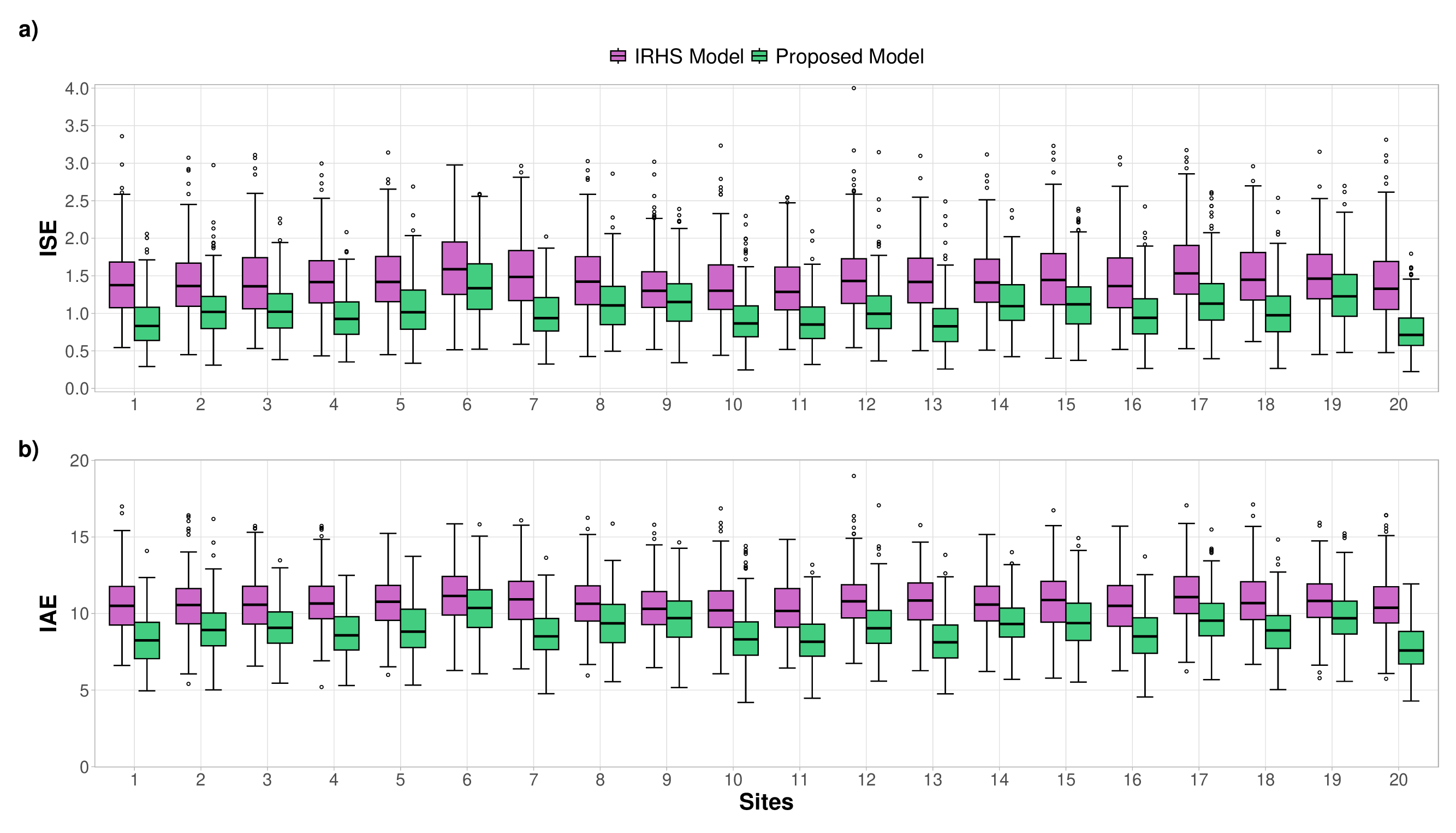}
	\caption{Reconstruction accuracy of the proposed model versus the IRHS baseline under the moderate-noise regime. (a) ISE and (b) IAE across all 20 locations over 300 MC simulations.}
	\label{fig:study2_boxplot}
\end{figure}

Figure \ref{fig:study2_boxplot} provides a detailed comparison of the reconstruction accuracy between the proposed hierarchical spatial model and the baseline IRHS model. This evaluation is performed in a moderate-noise regime, a critical threshold at which local high-frequency features begin to merge with observational noise. Panel (a) shows the distributions of the Integrated Squared Error (ISE), while Panel (b) depicts the Integrated Absolute Error (IAE) across all 20 spatial locations. Both ISE and IAE serve as formal discrepancy metrics that quantify global estimation error over the entire functional domain; a detailed theoretical formulation of these discrepancy measures is provided in Appendix \ref{sec:app_metrics}. 

The results clearly demonstrate the empirical advantage of explicitly incorporating spatial dependence. In both evaluation metrics, the proposed spatial model (green boxplots) consistently shows lower median errors and significantly narrower interquartile ranges than the IRHS baseline (purple boxplots) across the entire spatial domain. The IRHS model, which limits inference by completely excluding the Matérn spatial correlation matrix and assuming strictly independent curves, tends to show systematically higher estimation errors and a greater vulnerability to extreme outliers. This performance gap underscores a fundamental limitation of independent estimators: without the ability to leverage neighboring locations, the independent continuous shrinkage prior struggles to accurately distinguish genuine local signal fluctuations from random background noise. By utilizing the underlying spatial topology through the Matérn correlation structure, the proposed model effectively shares information across the network. This spatial framework stabilizes parameter estimation, significantly reducing both squared and absolute reconstruction errors, and prevents the decline in statistical quality that commonly occurs when spatial dependence is ignored.
\subsection{Simulation Study 3: Predictive Performance}\label{sec:sim3} 
In this third simulation study, we evaluate the proposed framework's predictive capabilities for estimating functional trajectories at unobserved geographic coordinates. Building upon the moderate-noise regime (Scenario 2, $\sigma_{\text{noise}}=0.30$) established in Section \ref{sec:sim1}, this analysis compares our methodology against two distinct baseline models: the classical OKFD approach \citep{giraldo:2011} and the hierarchical Bayesian wavelet model proposed by \citet{song:2019}.

To facilitate a rigorous and equitable methodological comparison within a unified computational environment, we reconstructed the \citet{song:2019} framework (hereafter S\&M) using Stan. As the original implementation was developed in \texttt{MATLAB} and, to the best of our knowledge, its source code is not publicly available, we carefully adapted the framework to reproduce their theoretical specifications. In particular, we assigned a distinct exponential spatial process to each wavelet coefficient and employed their Laplace spatial prior configuration. Implementing both Bayesian approaches in \texttt{Stan} ensures that any observed differences in estimation accuracy or execution time are attributable solely to the underlying statistical architectures, rather than to discrepancies in programming languages or sampling algorithms. The parameter inflation inherent to the \citet{song:2019} model necessitates computing and Cholesky-factorizing distinct spatial covariance matrices for every coefficient at each leapfrog step of the NUTS, rendering evaluation of the full set of 300 Monte Carlo replicates computationally prohibitive. Therefore, we conducted the predictive assessment using the first 100 independent replicates from Scenario 2. This subset provides a robust, statistically sufficient sample size to characterize the distribution of predictive errors and assess MCMC efficiency, preserving the rigor of the comparative analysis while remaining computationally feasible.

As illustrated in the spatial grid in Panel (a) of Figure \ref{fig:study1_data}, spatial locations 9 and 19 were intentionally excluded from the model training phase. This exclusion allows for the evaluation of predictive performance under different topological constraints. Site 9 is situated in a central area with a high density of neighboring stations, while Site 19 is located at the periphery of the domain and is geographically isolated. These two locations serve as new, unobserved points, requiring the models to predict their functional trajectories based solely on geographic coordinates and the observed data from neighboring stations.

\begin{figure}[!hbt]	
	\centering
	\includegraphics[width=1\textwidth]{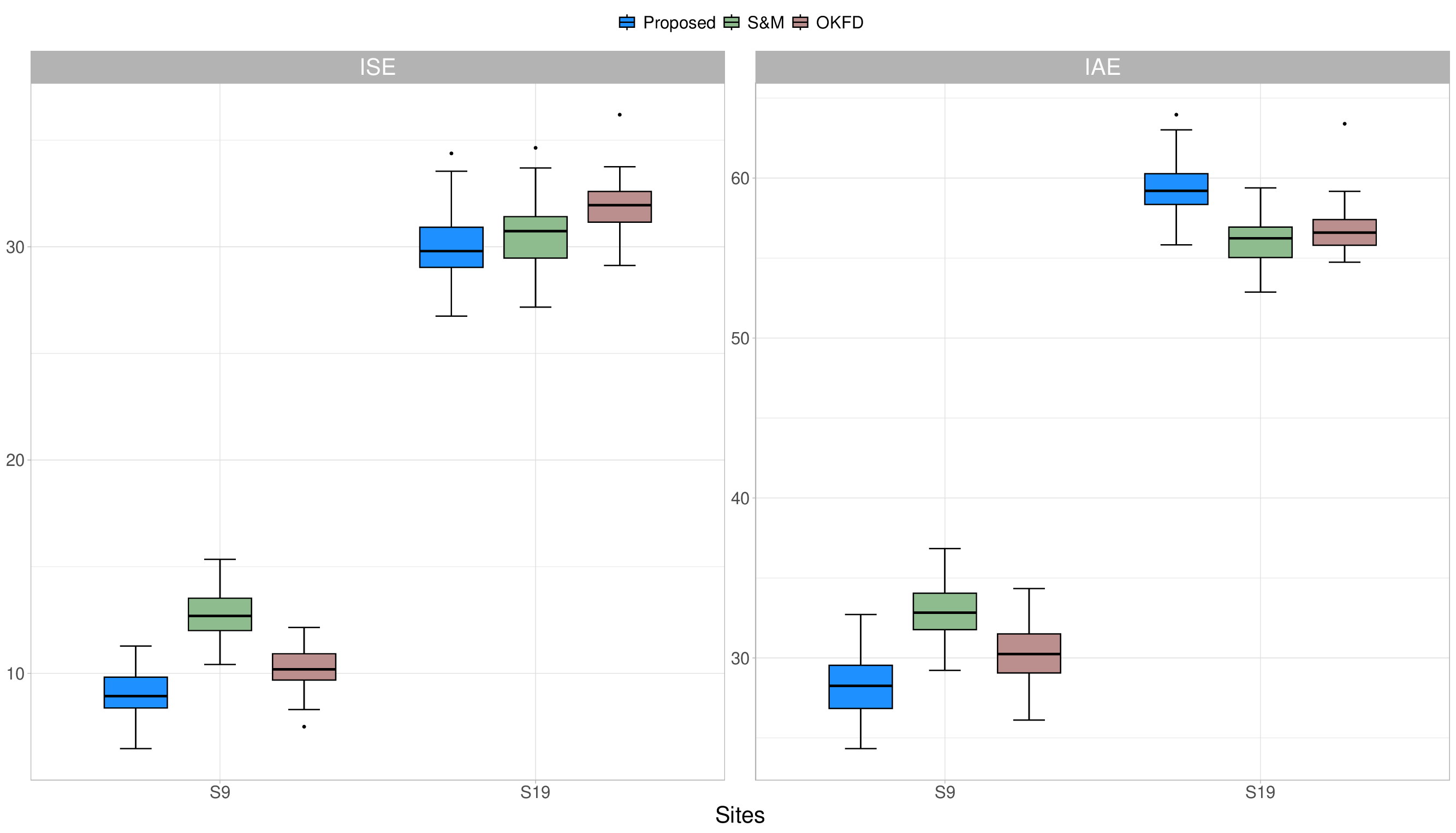}
	\caption{Predictive error at the held-out sites S$_9$ (central) and S$_{19}$ (peripheral) over 100 Monte Carlo replicates of Scenario 2 \(\sigma_{\text{noise}} = 0.30\). Panels show the distributions of the ISE and the IAE for the proposed model, the \citet{song:2019} model (S\&M), and the OKFD.}
	\label{fig:Simulation Study 3}
\end{figure}

Figure \ref{fig:Simulation Study 3} and Table \ref{tab:predictive-comparison} jointly summarize the assessment. Figure \ref{fig:Simulation Study 3} presents the full ISE and IAE distributions for the three competing models. At the same time, Table \ref{tab:predictive-comparison} reports the corresponding means (MISE, MIAE) and, for the two Bayesian models, the empirical coverage and mean width of the pointwise 95\% HPD intervals. Since OKFD provides only point predictions, the interval-based comparison is limited to the proposed model and S\&M. Predictive error at the held-out sites S$_9$ (central) and S$_{19}$ (peripheral) is evaluated over 100 MC replicates of Scenario 2. The panels display the distributions of the ISE and IAE for the proposed model, the S\&M model, and OKFD. Point-prediction accuracy depends strongly on the density of neighboring stations around the target site. At the central site S$_9$, the proposed model achieves the lowest and most concentrated error distributions in Figure \ref{fig:Simulation Study 3}, a result confirmed numerically. Compared to S\&M, it reduces the MISE from 12.75 to 9.04 and the MIAE from 33.05 to 28.30. When neighboring stations are dense, the model benefits from abundant spatial information, which enhances the reconstruction of localized, high-frequency features.

At the isolated peripheral site S$_{19}$, the three error distributions overlap substantially, and the ranking becomes far less clear-cut. The proposed model retains a marginal edge on the squared error metric (MISE 29.96 versus 30.55), whereas S\&M attains a slightly lower absolute error (MIAE 56.04 versus 59.36). This apparent reversal follows from what each metric rewards. The ISE weights deviations quadratically and is therefore dominated by the few sharp peaks that the wavelet basis recovers faithfully. Still, the smoother competitors flatten, keeping the proposed model ahead on MISE. The IAE weights every point in proportion to its error, so those flattened peaks contribute only linearly; there, with little neighboring information to exploit, the proposed model leaves a small, diffuse error that accumulates over the long smooth stretches and slightly exceeds the concentrated peak error of the smoother methods. This near-parity is consistent with the edge effects observed elsewhere in the study: an isolated site offers little spatial information to any method, which narrows the gap between competitors.

The primary distinction between the two Bayesian models lies in the quality of their predictive intervals. At both sites, the proposed model achieves an empirical coverage of 0.97, which is close to the nominal value of 0.95, while generating substantially narrower intervals, with mean widths of 1.01 at S$_9$ and 1.47 at S$_{19}$. In comparison, the S\&M model attains a coverage of 1.00 at both sites, but only by producing intervals approximately twice as wide (2.06 and 2.82). Thus, the S\&M model achieves higher coverage at the expense of informativeness, whereas the proposed model maintains coverage near the nominal level with considerably more precise intervals. 
\begin{table}[htbp]
\centering
\caption{Predictive performance at the held-out sites S$_9$ and S$_{19}$: mean
integrated squared error (MISE), mean integrated absolute error (MIAE),
empirical coverage of the pointwise 95\% HPD intervals, and their mean length,
for the proposed model and for S\&M.}
\label{tab:predictive-comparison}
\begin{tabular}{llcccc}
\toprule
Site & Method & MISE & MIAE & 95\% HPD coverage & Mean length \\
\midrule
\multirow{2}{*}{S$_9$}    & Proposed   &  9.04 & 28.30 & 0.97 & 1.01 \\
                          & S\&M       & 12.75 & 33.05 & 1.00 & 2.06 \\
\midrule
\multirow{2}{*}{S$_{19}$} & Proposed   & 29.96 & 59.36 & 0.97 & 1.47 \\
                          & S\&M       & 30.55 & 56.04 & 1.00 & 2.82 \\
\bottomrule
\end{tabular}
\end{table}
\section{Real Data Application}\label{sec:realdata}
Interactions between geographic conditions and seasonal weather cycles directly influence atmospheric particulate matter (PM$_{10}$) concentrations. In the Valley of Mexico, the high altitude and mountainous terrain often hinder pollutant dispersion. While temperature and intense solar radiation at tropical latitudes catalyze atmospheric chemical reactions, the lack of precipitation and a stable mixing layer are the main drivers of the accumulation of suspended solid and liquid particles. Excess PM$_{10}$ poses a severe risk to respiratory and cardiovascular health, a concern that intensifies during the transition from the dry season to the rainy season.
\begin{figure}[!hbt]	
	\centering
	\includegraphics[width=1\textwidth]{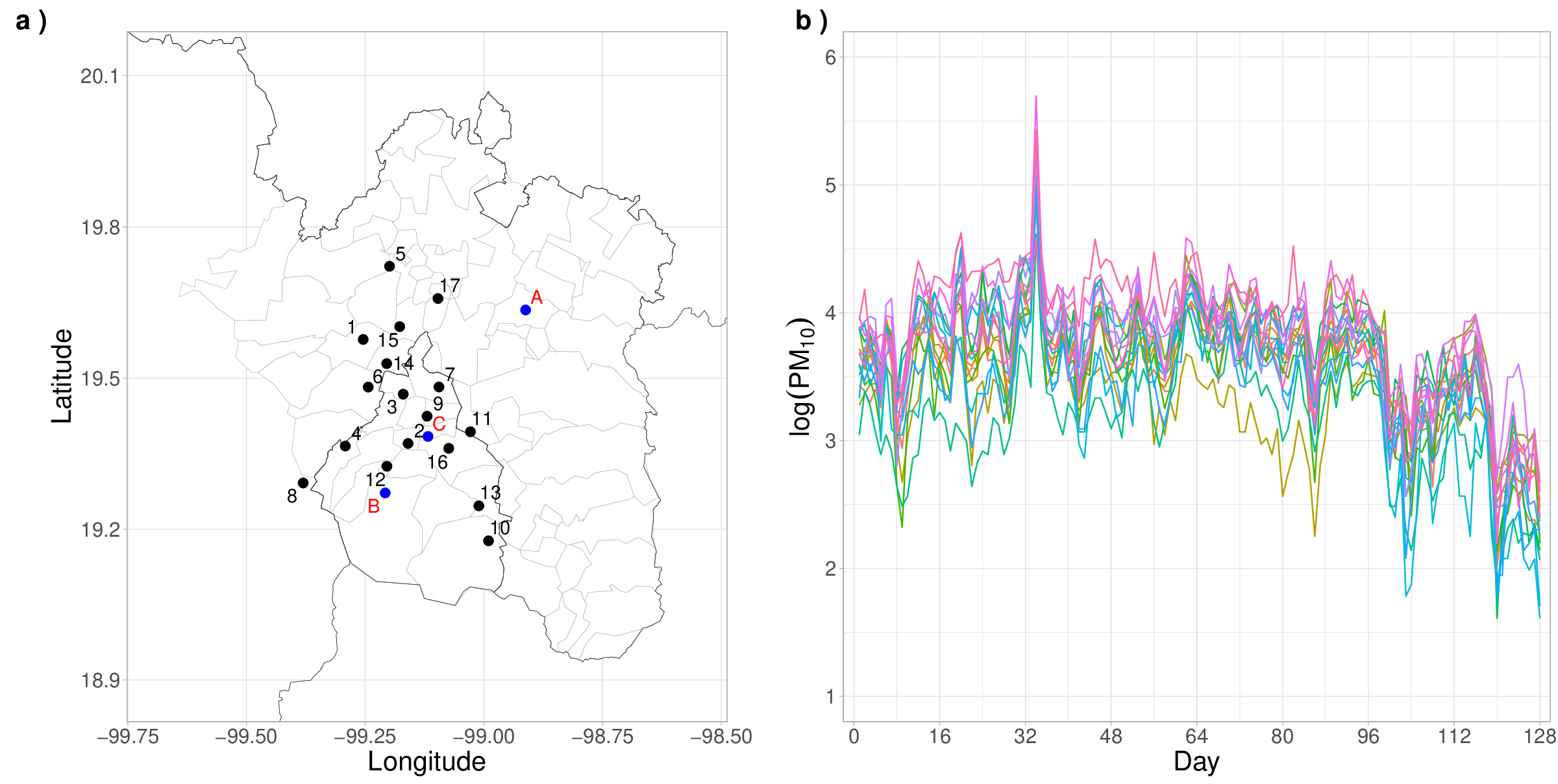}
	\caption{(a) Monitoring stations in Mexico City, highlighting the three held-out stations (A, B, and C) excluded from training for cross-validation. (b) Daily $\log(\text{PM}_{10})$ profiles for the monitored stations over 128 days from February 15 to June 22, 2025.}
	\label{fig:mexico_city_data}
\end{figure}

This study aims to assess PM$_{10}$ concentrations (measured in micrograms per cubic meter, $\mu g/m^3$) in Mexico City during a critical period of atmospheric transition in 2025. The Atmospheric Monitoring Network (RAMA - Red Automática de Monitoreo Atmosférico) collected the data using strategically located stations to capture air quality variability across the metropolitan area. The dataset includes consecutive hourly readings from early morning on February 15, 2025, through the end of the day on June 22, 2025. This timeframe enables observation of the pollutant's behavior under conditions of extreme aridity and the initial effects of the first convective summer rains. The raw data are publicly accessible on Mexico City's air quality portal at \url{http://www.aire.cdmx.gob.mx}. Given that environmental records often contain technical errors or maintenance periods (missing values), we needed to address gaps in the PM$_{10}$ data to create a continuous time series. To accomplish this, we used the \texttt{missForest} package in \texttt{R} for data imputation. This non-parametric method, based on random forests, is particularly effective for atmospheric time series because it captures nonlinear interactions among variables without assuming rigid distributional forms. We implemented the algorithm with 100 trees and five iterations to ensure stable reconstructed values.

Each time series consists of 3,072 equally spaced observations, which correspond to 128 units in the daily time domain. To create a daily-level aggregated representation, each series is divided into 128 blocks, each containing 24 consecutive observations. The median is used as the measure of central tendency for each block, rather than the mean, because it is more robust against outliers and better accounts for potential deviations from normality in the distribution of PM$_{10}$ concentrations. As a result of this procedure, each series is represented by a vector of 128 daily PM$_{10}$ values, corresponding to the period from February 15 to June 22.

To assess predictive performance in a real-world setting, we conducted spatial hold-out cross-validation. Three stations, labeled A, B, and C in Figure \ref{fig:mexico_city_data}, Panel a), were entirely withheld from the training set to probe distinct geographic regimes. Station C is located in a densely instrumented, central region and serves as a test of interpolation. Station A is situated in a comparatively isolated region and evaluates performance under sparse-neighborhood support. Station B, positioned on the southeastern boundary of the network, tests stability under extrapolation. At these three coordinates, the proposed wavelet model was compared against two baselines: the classical OKFD method \citep{giraldo:2011} and the hierarchical Bayesian wavelet model of S\&M, re-implemented in \texttt{Stan} as described in Section~\ref{sec:sim3}. Because only the two Bayesian models yield full posterior predictive distributions, we assess uncertainty quantification, empirical coverage, and the mean width of the 95\% HPD intervals for the proposed model and S\&M. In contrast, we compare OKFD solely on point-prediction accuracy.
\begin{figure}[!hbt]	
	\centering
	\includegraphics[width=1\textwidth]{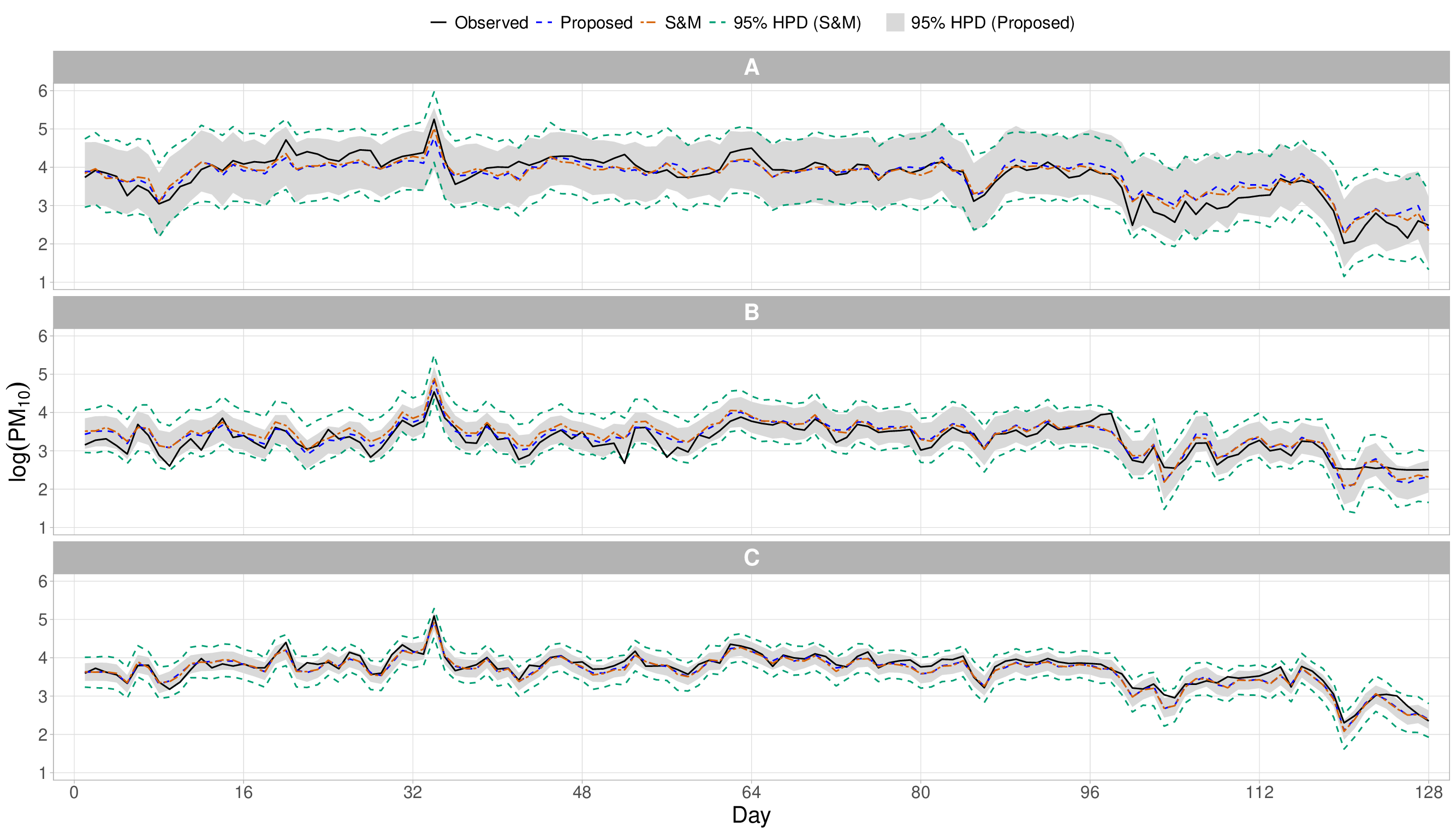}
	\caption{Predictive performance and uncertainty quantification at the three held-out stations (A, B, and C), comparing the proposed model with S\&M. Solid black: observed log-transformed PM$_{10}$. Dashed blue: proposed posterior predictive mean, with its 95\% HPD interval as the gray shaded region. Two-dash orange: S\&M posterior predictive mean, with its 95\% HPD interval bounded by dashed green lines.}
	\label{fig:pred_curves}
\end{figure}

Figure \ref{fig:pred_curves} presents a comparison of the posterior predictive reconstructions generated by the proposed model and S\&M at the three held-out stations. Both Bayesian models effectively capture the highly non-stationary and localized dynamics of the series, including the acute pollution episode observed around day 32. Their 95\% HPD bands encompass nearly all held-out observations, aligning with the empirical coverage reported in Table \ref{tab:error_comparison}. However, the two models differ significantly in the precision of their uncertainty estimates. The HPD band of the proposed model is consistently narrower than that of S\&M at each station, particularly at the central station C and the boundary station B, where the S\&M intervals are notably wider. Consequently, the proposed model achieves valid coverage while providing substantially more informative predictive intervals.
\begin{figure}[!hbt]	
	\centering
	\includegraphics[width=1\textwidth]{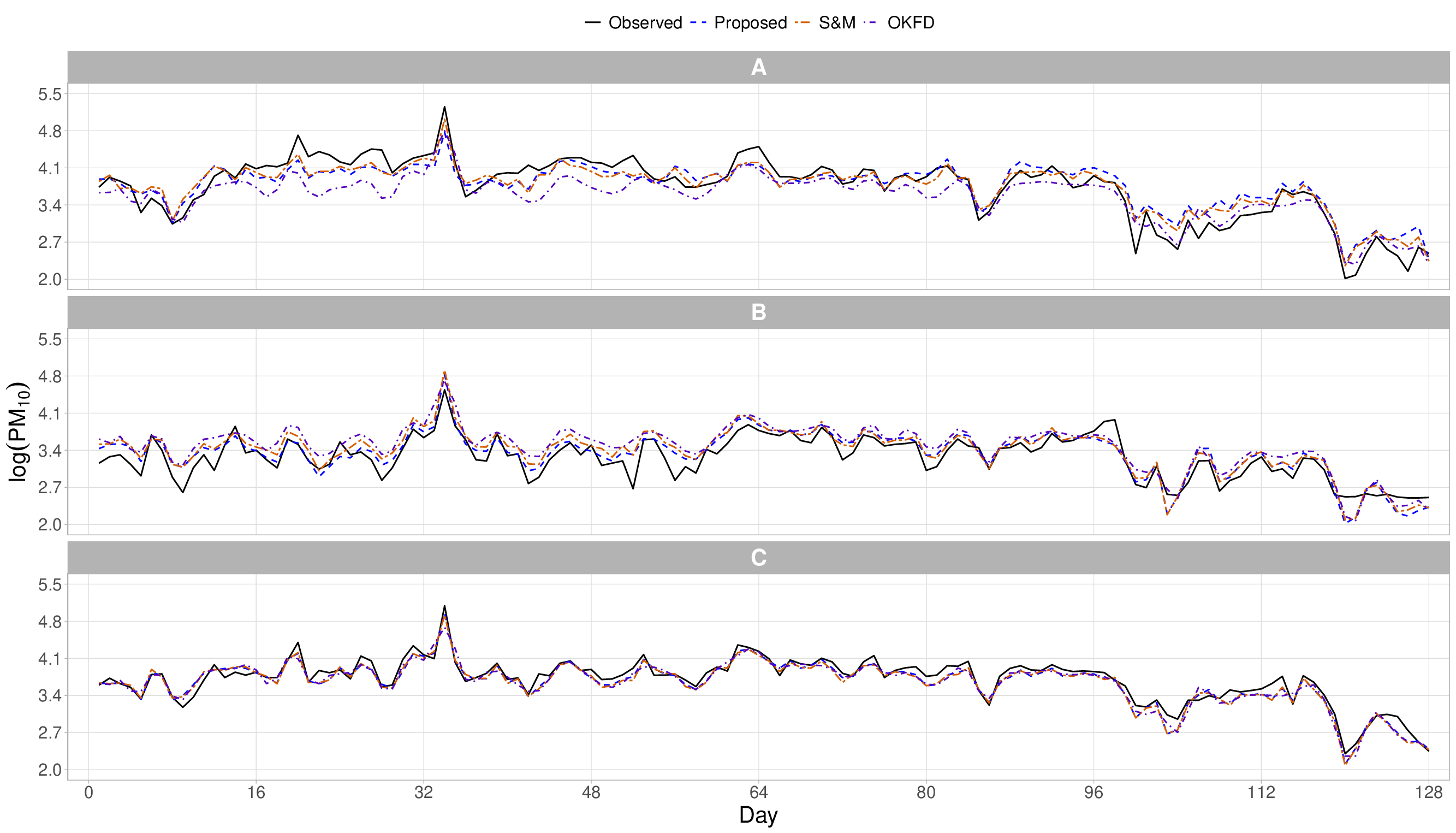}
	\caption{Point estimates at the held-out stations (A, B, and C). Solid black: observed $\log(\text{PM}_{10})$. Dashed blue: proposed model. Two-dash orange: S\&M. Dotdash: OKFD.}
	\label{fig:pred_curves2}
\end{figure}

Figure \ref{fig:pred_curves2} compares the point predictions of the three methods against the observed curves. OKFD, which relies on a penalized B-spline basis, over-smooths the trajectories and systematically underestimates the sharp, high-frequency peaks that characterize daily air-quality dynamics. Both wavelet-based models, by contrast, resolve these localized features far more faithfully: the multiresolution wavelet basis, coupled with the spatially informed Regularized Horseshoe prior, allows the proposed model to reconstruct smooth global trends and abrupt local spikes within a single adaptive framework.
\begin{table}[htpb]
\centering
\caption{Predictive performance at the three held-out stations (A, B, and C). ISE and IAE of the point predictions are reported for each method; empirical coverage and mean width of the 95\% HPD intervals are additionally reported for the two Bayesian models (the proposed model and S\&M). Bold indicates the best value across the three methods for each metric.}
\label{tab:error_comparison}
\begin{tabular}{llcccc}
\toprule
Station & Method & ISE & IAE & Coverage & Mean width \\
\midrule
\multirow{3}{*}{A}
 & Proposed          & 7.66  & 25.46 & 1.00 & 1.60 \\
 & S\&M              & \textbf{5.66} & \textbf{21.42} & 1.00 & 1.84 \\
 & OKFD              & 12.47 & 31.51 & \pending{OKFD} & \pending{OKFD} \\
\midrule
\multirow{3}{*}{B}
 & Proposed          & \textbf{4.70} & \textbf{19.04} & \textbf{0.96} & \textbf{0.81} \\
 & S\&M & 5.95 & 22.75 & 0.98 & 1.14 \\
 & OKFD              & 10.50 & 31.08 & \pending{OKFD} & \pending{OKFD} \\
\midrule
\multirow{3}{*}{C}
 & Proposed          & \textbf{1.87} & \textbf{12.24} & \textbf{0.95} & \textbf{0.47} \\
 & S\&M & 2.05 & 12.97 & 1.00 & 0.79 \\
 & OKFD              & 2.67 & 15.07 & \pending{OKFD} & \pending{OKFD} \\
\bottomrule
\end{tabular}
\end{table}

Table \ref{tab:error_comparison} presents the quantitative assessment. Compared to OKFD, the proposed model reduces both the ISE and the IAE at all three stations. Specifically, the ISE decreases from 12.47 to 7.66 at Station A, from 10.50 to 4.70 at Station B, and from 2.67 to 1.87 at Station C. Additionally, unlike OKFD, the proposed model provides calibrated predictive intervals. In comparison with S\&M, the results are more nuanced. At the central (C) and boundary (B) stations, the proposed model achieves lower ISE and IAE values. However, at the isolated Station A, the S\&M model demonstrates superior point accuracy (ISE 5.66 versus 7.66; IAE 21.42 versus 25.46). This outcome aligns with the anticipated reduction in the benefit of level-wise strength borrowing under sparse-neighborhood support, as previously observed in the simulation study. The proposed model consistently outperforms S\&M in uncertainty quantification: its 95\% HPD intervals are narrower at every station (1.60 versus 1.84 at A, 0.81 versus 1.14 at B, and 0.47 versus 0.79 at C), and its empirical coverage remains at or above the nominal 0.95, aligning more closely with the nominal level at Stations B and C (0.96 and 0.95, compared to 0.98 and 1.00 for S\&M, which tends to over-cover). In summary, the proposed model outperforms OKFD across all stations and metrics, matches or exceeds S\&M except in point accuracy at the most isolated station, and consistently provides sharper and better-calibrated predictive uncertainty. This combination of accuracy and well-quantified uncertainty is particularly valuable for prediction at unobserved environmental-monitoring locations.
\section{Conclusions}\label{sec:5}
This study introduces a Wavelet-Based Bayesian Hierarchical Model for modeling spatially correlated functional data. To address the computational limitations and parameter inflation found in previous approaches, we developed a parsimony-driven architecture that utilizes a single Matérn spatial process at each resolution level. By integrating this spatial structure with the RHS continuous shrinkage prior and employing a non-centered parameterization, our model maintains computational feasibility and inference stability through the NUTS algorithm in \texttt{Stan}.

Results from both simulated and real data demonstrate the methodology's consistent performance. In the simulation study, the model's ability to leverage spatial structure enabled accurate recovery of the true latent signal. The methodology proved robust even under extreme noise (low signal-to-noise ratio), maintaining control over the RISE and demonstrating superior overall accuracy compared to a reference model that assumes spatial independence. When applied to real PM$_{10}$ pollution data in Mexico City, the model showed significant advantages over the classical OKFD method. Using wavelet bases helped avoid the excessive smoothing typically associated with spline-based approaches, enabling adaptive reconstruction of acute pollution spikes at unobserved locations and providing well-calibrated functional credible intervals.

While the methodology offers a robust, computationally efficient alternative, this article does not cover certain aspects, leaving significant opportunities for future research. Firstly, the model's spatial structure relies on an isotropic Matérn correlation function. In environmental contexts with complex topography, such as mountain ranges or deep valleys, using anisotropic covariance functions or non-stationary spatial fields can provide greater flexibility in capturing directional spatial decay. Secondly, the use of the DWT imposes significant structural constraints, requiring the functional data to be organized on a regular grid and to have a length of \( n = 2^L \). In this study, we addressed these requirements through random forest imputation (missForest) and symmetric filling strategies to mitigate edge effects. However, a promising avenue for future research could involve developing models that natively handle curves with non-equidistant sampling. As discussed by \citet{song:2019}, integrating the lifting wavelet transform, which does not require regularly spaced samples, into our Bayesian framework could be beneficial. Finally, this study focused on univariate functional trajectories. A logical next step, with significant practical implications, would be to extend the hierarchical structure to a multivariate context. This would enable the joint modeling of multiple air pollutants or correlated climate variables. Such an advancement would facilitate the exploration of interdependencies among functional processes, thereby enhancing the methodology's predictive power and applicability in complex environmental monitoring networks.
\subsubsection*{Acknowledgments}
This study was financed by the São Paulo Research Foundation (FAPESP), Brazil. Process numbers 2023/18036-3 and 2023/02538-0.
\subsubsection*{Conflict of interest}
The authors declare no potential conflict of interest.
\subsubsection*{Data Availability Statement}\label{data-availability-statement}
The R code used for simulations in Section \ref{sec:3} and the analysis of real data in Section \ref{sec:realdata} are available at \url{https://github.com/aaburbano/SFDA-Horseshoe-Wavelets}.
\bibliographystyle{plainnat}
\bibliography{Refe}
\appendix
\section{Measures of Discrepancy} \label{sec:app_metrics}
Estimating a function at a point is analogous to estimating a scalar parameter, and expectations ($\mathbb{E}[\cdot]$) and variances ($\mathbb{V}[\cdot]$) are computed with respect to the (unknown) distribution of the associated random variable, in practice approximated by the empirical distribution. For estimation at a single point $t$, a common metric is the Mean Squared Error (MSE), \(\text{MSE}[\widehat{Y}(t)] = \mathbb{E}[(\widehat{Y}(t) - Y(t))^2] = [\mathbb{E}[\widehat{Y}(t)] - Y(t)]^2 + \mathbb{V}[\widehat{Y}(t)]\). 

To evaluate the estimator over the entire domain rather than at a fixed point, a global criterion measuring the distance between $\widehat{Y}(t)$ and $Y(t)$ is required. The $L^{p}$ norm \citep{wand1994, gentle2009} of the error is given by \(\left(\displaystyle\int_{T} |\widehat{Y}(t) - Y(t)|^p dt\right)^{1/p}\), where $T$ is the domain of the true function $Y$, over which $\widehat{Y}(t)$ must also be defined, and the integral need not always exist. The $L^{2}$ norm yields the integrated squared error (ISE), \(\text{ISE}(\widehat{Y}(t)) = \displaystyle\int_{T} (\widehat{Y}(t) - Y(t))^2 dt\), and the $L^{1}$ norm is the integrated absolute error (IAE), \(\text{IAE}(\widehat{Y}(t))=\displaystyle\int_{T} |\widehat{Y}(t) - Y(t)| dt\).

The latter is invariant under monotonic transformations of the coordinate axes, unlike the $L^2$ norm. The magnitude of the ISE is largely influenced by the scale and total energy of the true function, which can be misleading when curves have varying amplitudes or when assessing performance across different SNRs. To obtain a scale-free metric, the Relative Integrated Squared Error (RISE) \citep{shamshoian:2025} standardizes the ISE by the total energy (the squared $L^{2}$ norm) of the true latent signal, \(\text{RISE}(\widehat{Y}(t)) = \displaystyle \int_{T}(\widehat{Y}(t)-Y(t))^{2}dt\Big/\displaystyle \int_{T}(Y(t))^{2}dt\).

Since the data are modeled on a discrete grid of $n$ points across $T$, these integrals are approximated numerically using the trapezoidal rule.
\end{document}